\documentclass[journal]{IEEEtran}

\input epsf
\usepackage{graphicx}
\usepackage{multirow}
\usepackage{cite}
\usepackage{amsmath,amssymb,amsfonts,amssymb}
\usepackage{graphicx}
\usepackage{textcomp}
\usepackage{xcolor}
\usepackage{mathrsfs}
\usepackage{array}
\allowdisplaybreaks 
\usepackage{graphicx}
\usepackage{caption}
\usepackage{verbatim}
\usepackage{enumitem}
\usepackage{booktabs}
\usepackage{mathabx}
\usepackage{setspace}
\usepackage{enumitem}
\usepackage[T1]{fontenc}
\usepackage[latin9]{inputenc}
\usepackage{amstext}
\usepackage{amssymb,amsmath}
\allowdisplaybreaks
\usepackage{graphicx}
\usepackage{caption}
\usepackage{subfigure}
\usepackage{verbatim} 
\usepackage[lined,ruled,commentsnumbered]{algorithm2e}
\usepackage{hyperref}

\hypersetup{
    colorlinks=true,
    linkcolor=black,
    filecolor=black,      
    urlcolor=black,
    citecolor=black,
}\usepackage[lined,ruled,commentsnumbered]{algorithm2e}
\makeatletter
\def\namedlabel#1#2{\begingroup
  #2%
  \def\@currentlabel{#2}%
  \phantomsection\label{#1}\endgroup
}


\newtheorem{theorem}{Theorem}

\newtheorem{lemma}{Lemma}

\IEEEoverridecommandlockouts

\begin{document}

\title{DeepTrace: Learning to Optimize Contact Tracing in Epidemic Networks with Graph Neural Networks}

\author{ Chee~Wei~Tan, Pei-Duo~Yu, Siya~Chen and~H. Vincent Poor 
\thanks{C. W. Tan is with Nanyang Technological University, Singapore, Nanyang Ave., Singapore (e-mail: cheewei.tan@ntu.edu.sg).}
\thanks{P. D. Yu is with the Department of Applied Mathematics, Chung Yuan Christian University, Taiwan (e-mail: peiduoyu@cycu.edu.tw).}
\thanks{S. Chen is with the Department of Computer Science, City University of Hong Kong, Hong Kong (e-mail: siyachen4-c@my.cityu.edu.hk).}
\thanks{H. V. Poor is with the Department of Electrical and Computer Engineering, Princeton University, Princeton, NJ 08544 USA (e-mail: poor@princeton.edu).}
\thanks{This work is supported in part by the U.S. National Science Foundation under RAPID Grant IIS-2026982, the National Science and Technology Council of Taiwan under Grant 112-2115-M-033-004-MY2, Hong Kong ITF Project ITS/188/20, Singapore Ministry of Education Academic Fund (RG91/22) and National Medical Research Council CS-IRG (CIRG24jul-0021), Sayling Wen Cultural \& Educational Foundation Faculty Scholarship, the Institute for Pure and Applied Mathematics fellowship, C3.ai Digital Transformation Institute and Princeton Language and Intelligence.}

}

\maketitle

\begin{abstract}
\textcolor{black}{Digital contact tracing aims to curb epidemics by identifying and mitigating public health emergencies through technology. Backward contact tracing, which tracks the sources of infection, proved crucial in places like Japan for identifying COVID-19 infections from superspreading events. This paper presents a novel perspective on digital contact tracing by modeling it as an online graph exploration problem, framing forward and backward tracing strategies as maximum-likelihood estimation tasks that leverage iterative sampling of epidemic network data. The challenge lies in the combinatorial complexity and rapid spread of infections. We introduce \textit{DeepTrace}, an algorithm based on a Graph Neural Network that iteratively updates its estimations as new contact tracing data is collected, learning to optimize the maximum likelihood estimation by utilizing topological features to accelerate learning and improve convergence. The contact tracing process combines either BFS or DFS to expand the network and trace the infection source, ensuring efficient real-time exploration. Additionally, the GNN model is fine-tuned through a two-phase approach: pre-training with synthetic networks to approximate likelihood probabilities and fine-tuning with high-quality data to refine the model. Using COVID-19 variant data, we illustrate that \textit{DeepTrace} surpasses current methods in identifying superspreaders, providing a robust basis for a scalable digital contact tracing strategy.
}
\end{abstract}
\IEEEoverridecommandlockouts

%
\IEEEpeerreviewmaketitle

\section{Introduction}
During the early days of the COVID-19 pandemic, a multitude of mobile applications \cite{landau2021people, bengio2020predicting,infocom2021, stevens2020tracetogether} were deployed to detect individuals potentially exposed to the SARS-CoV-2 coronavirus and mitigate its transmission. Digital contact tracing, driven by network analytics and machine learning, continues to play a critical role in epidemic surveillance and preventing hospital-acquired infections by optimizing coverage and tracking the spread of infectious diseases \cite{tanbigdata}. Its primary objective is to curb the transmission of epidemics, but this is complicated by challenges such as rapid disease spread, asymptomatic carriers, and the disproportionate impact of superspreaders (e.g., approximately $20\%$ of infected individuals account for $80\%$ of transmissions \cite{adam2020clustering}). The development of robust, automated digital contact tracing systems leveraging network science and machine learning remains in its early stages.

It has been recently shown in \cite{cencetti2021digital, kojaku2021effectiveness, bradshaw2021bidirectional,bengio2020predicting} that compared with `forward' tracing (i.e., to whom disease spreads), `backward' tracing (from whom disease spreads) can be more effective due to overlooked biases arising from the heterogeneity in contacts. Tracing sources of spreading (i.e., backward contact tracing), as had been used in Japan \cite{kojaku2021effectiveness}, has proven effective as going backward can pick up infections that might otherwise be missed at superspreading events. The ability to identify the source of spreading (i.e., the Patient Zero or superspreaders in a pandemic) is essential to contact tracing. In this paper, we propose a novel formulation of the contact tracing problem that integrates both forward and backward tracing, utilizing maximum likelihood estimation as a subproblem. Our approach focuses on {\it optimizing who to track and how to track} through graph neural network learning, employing network centrality metrics to design fast and efficient contact tracing algorithms.

\subsection{Related Works} \label{sec:related_works}
The contagion source inference problem was first studied in the seminal work \cite{ShahTransIT2011} as a maximum likelihood estimation problem based on a Susceptible-Infectious (SI) model \cite{bailey1975mathematical}, which is a special case of the classical Susceptible-Infected-Recover (SIR) model in epidemiology \cite{grassly2008mathematical,chen2020time, eletreby2020effects}. This problem becomes challenging with large graphs, as the scale complicates tracking spreading likelihoods and calculating permissible permutations for each node. However, the degree-regular tree special case can be solved in polynomial time using a network centrality approach known as rumor centrality \cite{ShahTransIT2011, shah2012rumor}, which is proportional to the number of permitted permutations. Equivalently, this optimal likelihood estimate is the tree centroid \cite{zelinka1968medians,JSTSP2018,nowbook2023}. The network centrality approach \cite{ShahTransIT2011} has been extended to scenarios like random trees \cite{Vassio2014, K2013}, multiple snapshot observations \cite{wang2014rumor}, multiple sources \cite{Luo2013} and the {\it epidemic centrality} for a disease pandemic in \cite{pd_jstsp2022}. Unlike the snapshot model \cite{ShahTransIT2011}, other studies \cite{sridhar2020sequential, meirom2015, sridhar2021bayes,Yu2024} use time series data, Bayesian analysis, and stochastic processes to solve the contagion source inference problem via sequential and quickest detection techniques. For a comprehensive overview of the work on detecting contagion sources, we refer the reader to \cite{nowbook2023}.

However, computing the ML estimator for the source on large epidemic networks can be computationally challenging \cite{ShahTransIT2011, pd_jstsp2022}, particularly as the network size grows and complexity increases with the presence of cycles. Thus, using Graph Neural Networks (GNNs) is a viable option. Various studies have used GNNs for pandemic control \cite{la2020epidemiological, shah2020finding, patient_zero_aaai, chen2022identifying}. Gatta \cite{la2020epidemiological} focuses on estimating epidemiological model parameters, differing from other works \cite{shah2020finding, patient_zero_aaai, chen2022identifying} that aim to identify patient zero (superspreaders). Shah \cite{shah2020finding} examines SIR and SEIR models with graph convolutional networks, but these are limited to transductive settings and unable to handle graph inputs of varying sizes. In \cite{patient_zero_aaai}, susceptible nodes were pre-labeled as unlikely sources. Meanwhile, \cite{patient_zero_aaai, shah2020finding} used the \emph{true} outbreak source as the label, which can diverge from the maximum-likelihood (ML) estimator. Using the ML estimator as the label, we align training with identifying the node most likely to explain the data, thus reducing bias and improving accuracy.

To address the issues related to the training set \cite{patient_zero_aaai,shah2020finding}, and the computational complexity challenges discussed in \cite{shah2012rumor,pd_jstsp2022}, we propose a GNN-based framework with a two-phase semi-supervised training approach. In this framework, we use graph data with the ML estimator as the superspreader during the fine-tuning phase. Specifically, we adopt the SI spreading model, as thoroughly studied in \cite{shah2012rumor,pd_jstsp2022,nowbook2023}. It has been established that the ML estimator is equivalent to certain network centers \cite{JSTSP2018, ShahTransIT2011,pd_jstsp2022}. This equivalence allows us to easily obtain high-quality labeled data for specific network topologies, thereby improving the accuracy of identifying the most likely superspreader.

Lastly, our previous work \cite{chen2022identifying} serves as a pilot study on using GNNs for backward contact tracing, framing contact tracing as online graph exploration. Compared to the conference version, we have studied the performance of the DFS strategy (cf. Lemma \ref{lem:comp_S} and Theorem \ref{thm:DFS_main}). Additionally, Theorem \ref{thm:thm3} demonstrates the feasibility of using GNNs to {\it learn to optimize} maximum likelihood estimators in graphs. In our experimental simulations, we have included numerous tests, evaluating the DFS and BFS strategies combined with various source detection methods from the literature across different network topologies.

\subsection{Main Contributions}
{\color{black}
\begin{itemize}
    \item We frame digital contact tracing as an online graph exploration problem, deriving both forward and backward tracing strategies from a maximum likelihood estimation that captures the pandemic's dynamic spread. To address the challenging nonconvex optimization involved, we employ modified graph neural networks to predict network centrality measures. Our low-complexity algorithm, DeepTrace, refines the prediction of the most likely contagion source as data is collected, effectively applying it to epidemic networks of various sizes.
    
    \item We utilize GNNs to perform the task of backward contact tracing. By training the GNN to learn to optimize the maximum likelihood estimation (MLE) problem, we enhance its ability to solve the complex nonconvex optimization task of identifying the contagion source. We integrate network centrality methods from the literature for calculating MLE for the source to generate high-quality training data, thereby improving the efficiency and accuracy of our GNN training.
    
    \item For the forward contact tracing component, we employed graph theory and algorithmic analysis to thoroughly examine the differences between BFS and DFS strategies. Our experimental simulations demonstrated that using BFS to expand the epidemic network enables faster detection of superspreaders compared to the DFS strategy.
    
    \item Using datasets from synthetic epidemic networks and COVID-19 contact tracing in Taiwan and Hong Kong, we show that Algorithm DeepTrace outperforms state-of-the-art heuristics from \cite{ShahTransIT2011,pd_jstsp2022}.
\end{itemize}
}


\section{Contact Tracing Problem Formulation}
Let us first introduce the spreading models, including one relevant to COVID-$19$ in \cite{eletreby2020effects}, and then present the digital contact tracing optimization problem formulation followed by its mathematical analysis.
\subsection{Epidemic Spreading Model} \label{sec:SI_Model}
There are numerous models in the literature in the study of the spread of infectious diseases ranging from commonly well-known ones like the Susceptible-Infectious model to the Susceptible-Infectious-Recovery-Susceptible (SIRS) models \cite{grassly2008mathematical, sridhar2020sequential, chen2020time, eletreby2020effects}. Despite being the most basic one, the SI model can accurately model a pandemic in its early stage (e.g., COVID-$19$ ca. 2020), whereas other SIRS models \cite{eletreby2020effects} incorporate complex factors like virus mutation, societal response (e.g., mask-wearing and social distancing) and the use of vaccines (e.g., COVID-$19$ ca. 2021).

We primarily focus on the SI model for our analysis, as the characteristics of the COVID-19 virus can persist in individuals for extended periods without a clear recovery or removal phase. Studies have shown that SARS-CoV-2 can remain detectable in patients for several weeks, even after symptoms subside \cite{duration2020}. This persistent detectability supports using the SI model in the early stage of the outbreak. The SI model provides a simplified yet relevant framework for our analytical purposes, focusing on the essential aspects of identifying the initial source of infection. The SI model assumes that there are only two types of individuals, susceptible and infectious, during the epidemic spreading. Also, we follow the assumption that the times between infection events are independent and exponentially distributed \cite{shah2012rumor} with an average infection time (with most estimates of COVID-$19$ incubation period ranging from 1 to 14 days \cite{eletreby2020effects}). The virus spreads in a social contact network can be modeled by a diffusion process on a graph denoted by $\mathbb{G}$. Let $V(\mathbb{G})$ and $E(\mathbb{G})$ denote the set of nodes and edges in $\mathbb{G}$, which represent people and their social contacts (an edge exists between two persons if, say, they are within 6 feet for more than 15 minutes) respectively. We shall call $\mathbb{G}$ the \textit{underlying network}, where the topology of $\mathbb{G}$ poses a constraint to the epidemic spreading \cite{ganesh2005,shah2012rumor,pd_jstsp2022}. 
The spreading of the virus starts from a node in $\mathbb{G}$ (i.e., the superspreader) and spreads to other nodes under the constraint that a susceptible node can be infected only if at least one of its neighboring nodes has been infected. At some point in time, all infected nodes in $\mathbb{G}$ form a connected subgraph of $\mathbb{G}$, denoted as $\mathbb{G}_N$ and is called the \textit{epidemic network}, where $N$ represents the number of nodes in the epidemic network. 

This epidemic network $\mathbb{G}_N$ and the infection spreading dynamics are assumed to be unknown to a contact tracer, and this is the chief uncertainty faced by all contact tracers, who therefore adopt a strategy to collect this data starting from an index case (i.e., the first documented infected person) and trace his or her close contacts (according to the topology of $\mathbb{G}$) and so on.

\subsection{Forward and Backward Contact Tracing}
\begin{figure*}
\centerline{\includegraphics[scale=0.42]{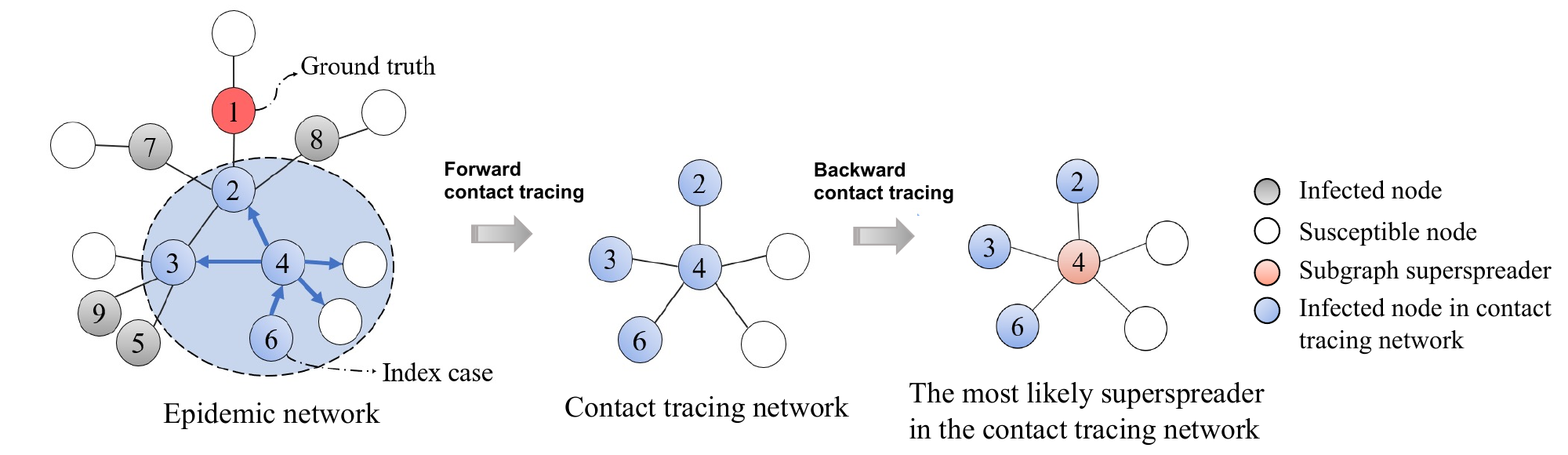}}
\captionsetup{font={footnotesize, stretch=1}, justification=raggedright}
\caption{\label{fig:fw_bw}Illustration of an epidemic network $\mathbb{G}_9$ with nine infections (shaded nodes) whose numbering indicates the infection order starts from the ground truth, i.e., the real superspreader. The contact tracing network $G_4$ (within a dotted circle) starts from the index case node $v_6$ (blue arrows show the tracing directions) by forward contact tracing. The backward contact tracing is to find the node in $\mathbb{G}_9$ that is most likely to be the superspreader.}
\end{figure*}

At various stages of contact tracing, we model an instantaneous snapshot of a subgraph of $\mathbb{G}_N$ that we call the {\it contact tracing network}. This contact tracing network grows by one (infected) node at each stage. If the traced node is infected, we continue to trace the node's neighbors; otherwise, we stop tracing along a path involving this node.  Let $G_{n}$ denote the contact tracing network with $n$ nodes at the $n$-th stage of tracing (thus, the index case is $G_1$). Hence we have $G_{n}\subseteq \mathbb{G}_N \subseteq \mathbb{G}$. A goal of contact tracing is to find the node in $\mathbb{G}_N$ that is most likely to be the superspreader, as shown in Fig. \ref{fig:fw_bw}. However, this most likely superspreader may not yet be in $G_{n}$, meaning that the contact tracing effort is still in its early stage or not fast enough (relative to the pandemic spreading speed). In such a case, backward contact tracing should yield an estimate as close as possible to this most likely superspreader. In other words, given the available data at the $n$-th stage, the contact tracer finds the node in $G_{n}$ that is the fewest number of hops away from the most likely superspreader in $\mathbb{G}_N$ (i.e., the optimal maximum likelihood estimate had this $\mathbb{G}_N$ been given entirely upfront to the contact tracer \cite{ShahTransIT2011}).

Given the data $G_{n}$ harvested by contact tracing at the $n$-th stage, we have the following maximum-likelihood estimation problem:
\begin{equation}
\hat{v} \in \arg \max_{v\in G_{n}\subseteq \mathbb{G}_N\subseteq \mathbb{G}} \mathbb{P}(G_n \mid v),
\label{eq:est}
\end{equation}
where $\mathbb{P}(G_{n} \mid v)$ is the likelihood function and $\hat{v}$ is the most-likely superspreader of the current observed $G_n$. The key challenge is that $\mathbb{G}_N$ is unknown to the contact tracer, who has to consider:

{\it Forward contact tracing}: How to construct the contact tracing network efficiently starting from a given index case? 

{\it Backward contact tracing}: How to solve \eqref{eq:est} to give the best instantaneous estimate of the superspreader given the data?


Answering both the forward and backward contact tracing jointly constitutes an iterative statistical inference process to track the most likely superspreader in the epidemic network. Specifically, for the forward contact tracing subproblem, it is natural to enlarge the forward contact tracing network with breadth-first search (BFS) or depth-first search (DFS) graph traversal starting from the index case. For the backward contact tracing subproblem, suppose that a given node $v$ in $G_{n}$ is the superspreader. Then, starting from this particular node, there are many possible ways to infect all the other nodes consistent with the given graph $G_{n}$ harvested by contact tracing at the $n$-th stage. Even though $\mathbb{G}_N$ is unknown, it is reasonable for a contact tracer to assume a possible infection order in $G_{n}$ by a {\it permitted permutation} given as $\sigma = \{ v_1=v, v_2,...,v_{n}\}$. Thus, we can reformulate problem (\ref{eq:est}) as:
\begin{equation}
\hat{v} \in \arg \max _{v\in G_{n}} \sum_{\sigma \in \Omega(G_{n} \mid v)} \mathbb{P}(\sigma | v),\label{eq:est1}
\end{equation}
where $\Omega(G_{n} \mid v)$ is the collection of all permitted permutations for $G_{n}$ rooted at $v$, and $\mathbb{P}(\sigma | v)$ is the probability of the permitted permutation $\sigma$ with $v$ as source.

Under the SI spreading model used in \cite{shah2012rumor,ShahTransIT2011,pd_jstsp2022}, the probability of a susceptible node being infected in the next time period is uniformly distributed among all outreaching edges from the infected nodes at the boundary of $G_{n}$, and the probability of a permitted permutation is as follows:

\begin{equation}
\label{eq:prob1}
\mathbb{P}(\sigma | v)
=\prod_{i=1}^{n-1} \frac{\Phi_i}{\sum_{j=1}^{i}[d(v_j)-2\Phi_{j-1}]},
\end{equation}
where $d(v_i)$ is the degree of node $v_i$ in $\mathbb{G}$, and $\Phi_i = |e(v_{i+1})\cap (\bigcup_{j=1}^{i}e(v_j))|$ with $e(v_i)$ being the set of edges connecting to node $v_i$. For the edge case, we define $\Phi_0=0$. The value of $\Phi_i$ counts the number of edges linked between node $v_{i+1}$ and its infected neighbors. Hence, the probability of being infected is proportional to the number of infected neighbors. Note that the probability defined in \eqref{eq:prob1} also works if $\mathbb{G}$ is a graph with multiple edges, i.e., the number of edges between a given pair of nodes can be more than one.

If $\mathbb{G}$ is a general tree as a special case, then every newly infected node, say $v_{i+1}$ only connected to one of the infected nodes; otherwise, $\mathbb{G}$ is not a tree. Hence, when $\mathbb{G}$ is a general tree, we have $\Phi_i=1$, for $i\neq 0$, and the probability of a permitted permutation can be given by \cite{ShahTransIT2011}: 
\begin{equation}
\label{eq:prob2}
\mathbb{P}(\sigma | v)
=\prod_{i=1}^{n-1} \frac{1}{\sum_{j=1}^{i}d(v_j)-2(i-1)}.
\end{equation}

\subsection{Regular Trees and General Graphs}
In this section, we will discuss the differences in using maximum likelihood estimation to find the source estimator when the underlying network is a regular tree versus a general graph. Additionally, we will explore how the known results for regular trees can be applied to general graphs. 

From \eqref{eq:prob1}, we can observe that when $\mathbb{G}$ is a regular tree, both $\Phi_i$ and $d(v_i)$ are constants which implies the value of \eqref{eq:prob1} is also a constant. Thus, the computation of the likelihood only depends on the value of $|\Omega(G_n, v)|$ in \eqref{eq:est1}, which simplifies the computation. On the other hand, it is still unknown whether there exists an efficient method for calculating the likelihood when $\mathbb{G}$ is a general graph since $\mathbb{P}(\sigma \mid v)$ is different for each possible $\sigma\in \Omega(G_n, v)$ and the size of $\Omega(G_n, v)$ grows exponentially with $n$. To overcome this computational complexity issue, we aim to leverage GNN to find the MLE. Various theoretical results on regular trees or unicyclic regular graphs \cite{ShahTransIT2011,pd_jstsp2022} provide us a way to generate the high-quality dataset; the label of each piece of data is the exact MLE of the epidemic graph for training GNN. The analysis in Theorem \ref{thm:bfs}, \ref{thm:DFS_main} shows the best and the worst possible outcomes during the contact tracing on regular trees, which provides us bounds on the performance of the contact tracing on general graphs.

Coming back to the contact tracer starts from the index case $G_1$ and collects more data in a forward manner (i.e., enlarging $G_{n}$ in \eqref{eq:est1}), the contact tracer also predicts the superspreader for that instant by solving (\ref{eq:est}). Intuitively, this means that as the contact tracing subgraph $G_{n}$ grows, the contact tracer desires this prediction to be closer (in terms of the number of hops in $\mathbb{G}_N$) to the most likely superspreader in $\mathbb{G}_N$. Considering this, how should $G_{n}$ grow in size in {\it forward contact tracing}? We propose to enlarge $G_n$ using BFS or DFS graph traversal \cite{awerbuch1987new, even2011graph}, which are detailed in the next section.

\section{Digital Contact Tracing by BFS and DFS}\label{sec:fw-tracing}
In this section, we detail interesting insights into tracing superspreaders in BFS tracing networks (using BFS traversal for forward contact tracing) and DFS tracing networks (using DFS traversal for forward contact tracing). Mathematically, the forward contact tracing process yields a rooted tree $\mathbb{G}_N$, whose nodes help solve \eqref{eq:est1}. For special cases, Theorem \ref{thm:bfs} states that this maximizer of \eqref{eq:est1}, i.e., the graph centroid of $\mathbb{G}_N$, converges to the graph centroid of $\mathbb{G}_N$ along a simple path on $\mathbb{G}_N$ when $N$ is fixed.

\subsection{Contact Tracing by Breadth-First Search}

\begin{figure}[h]
\centerline{\includegraphics[scale=0.19]{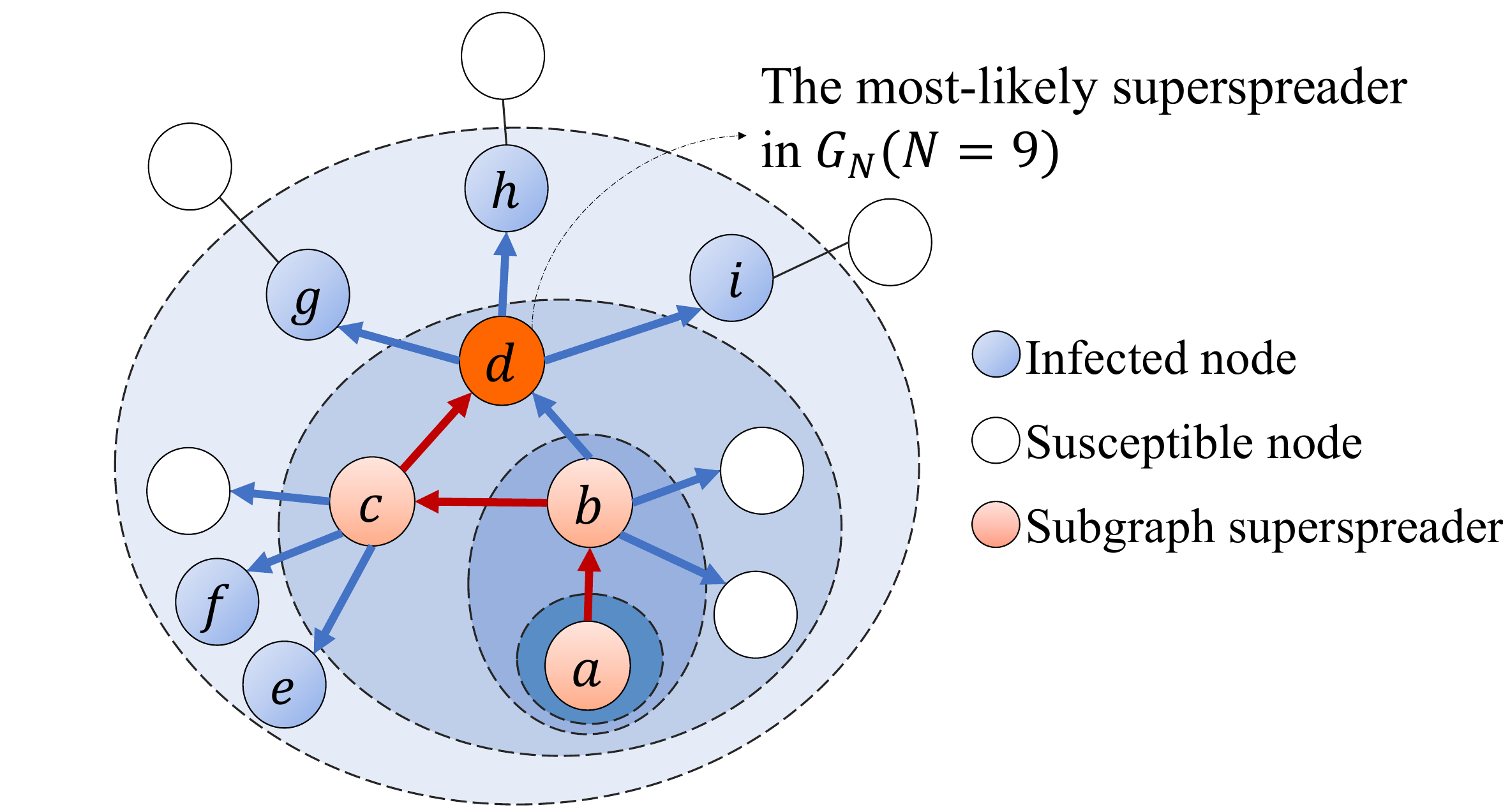}}
\captionsetup{font={footnotesize, stretch=1}, justification=raggedright}
\caption{\label{fig:maze}As the contact tracing network enlarges starting from the index case $v_a$ with BFS traversal, as ordered alphabetically $\{a,b,c,\ldots ,i\}$, the most likely superspreader given by \eqref{eq:est1} moves closer (in terms of number of hops) to the most-likely superspreader $v_d$ in the epidemic network $\mathbb{G}_N$ (indicated by red arrows).}
\end{figure}

Firstly, we propose to enlarge $\mathbb{G}_N$ using the BFS graph traversal algorithm \cite{awerbuch1987new, even2011graph}.
Fig. \ref{fig:maze} illustrates this with a tracer who starts from node $v_a$ and then moves to node $v_d$ (the most likely superspreader) gradually. 

Let $\textup{dist}(u,v)$ denote the graph distance (i.e., number of hops) between node $u$ and node $v$. If $N$ is fixed, then we have the following result showing that $\textup{dist}(v_{n}^*,v^*_N)$ is a decreasing sequence as $n$ grows when $\mathbb{G}_N$ is a $d$-regular tree, where $d$-regular tree is a tree that all non-leaf nodes have $d$ neighbors.

\begin{theorem}
\label{thm:bfs}
If $\mathbb{G}$ is a $d$-regular tree, and there exists a node $\overline{v}$ such that for any two leaf nodes $v_{\textup{leaf}}$ and $u_{\textup{leaf}}$ in $\mathbb{G}_N$, 
\begin{equation}
|\textup{dist}(v_{\textup{leaf}},\overline{v})-\textup{dist}(u_{\textup{leaf}},\overline{v})|\leq 1,\label{eq:condt}
\end{equation}
then the trajectory of $v_{n}^*$ is exactly the shortest path from the index case $v^*_1 $ to $v^*_N$ in $\mathbb{G}_N$.
\end{theorem}

Theorem \ref{thm:bfs} implies that in the special case where $\mathbb{G}$ is a $d$-regular tree, the epidemic network $\mathbb{G}_N$ satisfies \eqref{eq:condt}. By employing the BFS contact tracing strategy, we can progressively approach the optimal solution $v^*_N$ with each expansion of $G_n$. Thus, Theorem \ref{thm:bfs} underscores the effectiveness of the BFS strategy in achieving optimal solutions. 

Next, we provide an example, a class of graph structure satisfying \eqref{eq:condt}. If $G_{n}$ is a complete $N$-ary tree, then $|\textup{dist}(v_{\textup{leaf}},v_{n}^*)-\textup{dist}(u_{\textup{leaf}},v_{n}^*)|\leq 1$ for every $v_{\textup{leaf}}$ in $G_{n}$. Then, from Theorem \ref{thm:bfs}, we have that the trajectory of $v_{n}^*$ is exactly the shortest path from $v^*_1$ to $v^*_N$ in $\mathbb{G}_N$.

\subsection{Contact Tracing by Depth-First Search}
\begin{figure}
\centerline{\includegraphics[scale=0.2]{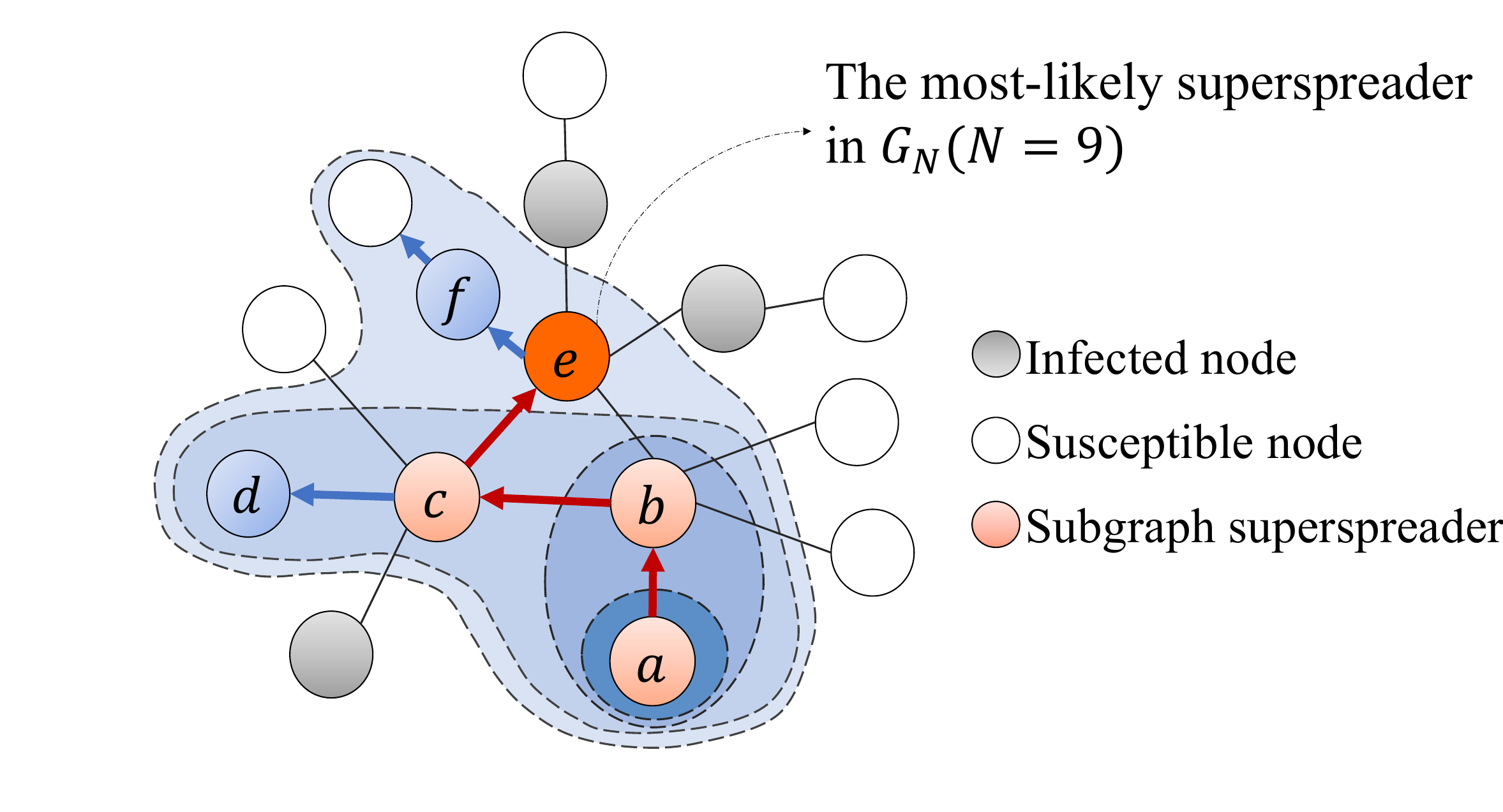}}
\captionsetup{font={footnotesize, stretch=1}, justification=raggedright}
\caption{\label{fig:dfs}As the contact tracing network enlarges starting from the index case $v_a$ with DFS traversal, as ordered alphabetically $\{a,b,c, \ldots ,f\}$, the most-likely superspreader given by \eqref{eq:est1} moves closer (in terms of number of hops) to the most-likely superspreader $v_e$ in the epidemic network $\mathbb{G}_N$ (indicated by red arrows).}
\end{figure}

Let us consider enlarging $\mathbb{G}_N$ with DFS tracing strategy \cite{even2011graph}, illustrated in Fig. \ref{fig:dfs}. We can categorize all consecutive pairs $(v_i^*,v_{i+1}^*)$ in $S$ into the following three types:
\begin{align*}
    &S_1 = \{(v_i^*,v_{i+1}^*)| v_i^*=v_{i+1}^*\},\\
    &S_2 = \{(v_i^*,v_{i+1}^*)| v_i^*\neq v_{i+1}^*,\;\textup{and}\;v_j^* \neq v_{i+1}^* ,\; \forall j<i\},\\
    &S_3 = \{(v_i^*,v_{i+1}^*)| v_i^*\neq v_{i+1}^* \;\textup{and}\;v_j^* = v_{i+1}^* \;\textup{for some}\; j<i\}.
\end{align*}

\def\svgwidth{120pt}   

The first type of consecutive pairs in $S_1$ represent the estimated superspreader that remains the same when the contact tracing network grows from $G_i$ to $G_{i+1}$. The second type, $S_2$ implies that the node $v_{i+1}^*$ is computed as the estimated superspreader for the first time. The third type, $S_3$  implies that $v_{i+1}^*$ had been chosen as the estimated superspreader before. Since the contact tracing network is a subgraph of a degree-regular tree, there are two maximum likelihood estimators when $\mathbb{G}_N$ can be divided into two subgraphs with the same size \cite{shah2012rumor}. We assume for each consecutive pair $(v_i^*,v_{i+1}^*)$ in $S_2$ and $S_3$, we have ${P}(G_{i+1} \mid v_{i+1}^*)>{P}(G_{i+1} \mid v_i^*)$, i.e., we only select $v_{i+1}^*$ as the new estimator when ${P}(G_{i+1} \mid v_{i+1}^*)>{P}(G_{i+1} \mid v_i^*)$. Let $|S_1|$, $|S_2|$, and $|S_3|$ denote the size of these three types of consecutive pairs, respectively. Let $T^v$ be a rooted tree of $\mathbb{G}_N$ rooted at $v$ and denote the subtree of $T^v$  rooted at $u$ as $T^v_u$. Note that the size of $S$ is $N$, so there are only $N-1$ consecutive pairs in $S$. Moreover, $S_1$,$S_2$ and $S_3$ are mutually exclusive. Hence, we can conclude that
\begin{equation}
    |S_1|+|S_2|+|S_3|=N-1.
\end{equation}

We introduce the following insights of the sequence $S$ by considering $S_1$, $S_2$, and $S_3$.

\begin{lemma}
\label{lem:comp_S}
Let $\mathbb{G}$ be a $d$-regular tree, and we apply the DFS tracing strategy, then we have $|S_1|\geq |S_2| \geq |S_3|$. Moreover, we can bound $|S_1|$, $|S_2|$ and $|S_3|$ as follows:
\begin{itemize}
    \item $|S_1|\geq (N-1)/2 \geq |S_2|$,
    \item $|S_3| \leq (N-1)/4$.
\end{itemize}
\end{lemma}
\textcolor{black}{
The edge cases of Lemma \ref{lem:comp_S} happen when $\mathbb{G}_N$ is a line graph with an odd number of nodes, and the index case is either the center or the leaf node of $\mathbb{G}_N$. We can deduce that more than half of those consecutive pairs in $S$ are type one, i.e., the estimated superspreader will remain the same most of the time. Lastly, this lemma serves as a tool when proving the following theorem.}

\begin{theorem}
\label{thm:DFS_main}
Let $\mathbb{G}$ be a $d$-regular tree and we apply DFS tracing strategy, if $v\in \mathbb{G}_N$ and $(\cdot ,v)\in S_3$ , then we have $| (\cdot ,v) | \leq \log_2(N)$, where $| (\cdot ,v) |$ denote the number of the pair $(\cdot ,v)$ in $S_3$.
\end{theorem}

Theorem \ref{thm:DFS_main} implies that when using the DFS tracing strategy to build the contact tracing network, each node in the epidemic network $\mathbb{G}_N$ should not be identified as the superspreader more than $\log_2(N)$ times as the contact tracing network expands iteratively. Next, let us compare the performance of both the DFS and BFS contact tracing strategies using two quantifiable metrics (that can be readily measured once the ground truth is given). The first metric is the average error, which is defined by
\begin{align}\label{eq:avg_err}
 \frac{1}{N}\sum\limits_{i=1}^{N}\textup{dist}(v^*_i,v^*_N).   
\end{align}

The second metric is the first detection time, which is the time when the estimated most-likely superspreader first matches the ground truth and is defined as
\begin{align}\label{eq:first_dect}
 \min  \{ i \; \vert \; v^*_i=v^*_N\}.   
\end{align}

The average error metric assesses the performance of any digital contact tracing algorithm, accounting for the selected forward contact tracing strategy and potential suboptimality in maximizing likelihood estimates. On the other hand, the first detection time measures how quickly the contact tracing algorithm can zoom in on identifying the most likely superspreader without a global view of the overall epidemic networks. It is desired that the most likely superspreader is the actual ground truth even as the epidemic network size $N$ increases independently of the digital contact tracing effort.

It is not surprising that the performance of any forward contact tracing strategy depends on the distance between the index case and the most likely superspreader, i.e., $\textup{dist}(v^*_1,v^*_N)$. For example, let the epidemic network $\mathbb{G}_{10}$ be a $3$-regular tree with six leaves. Assume that the index case is a leaf node, then the pair $(\textup{average error},\textup{first detection time})$ is $(0.7,6)$ and $(0.8,7)$ for the DFS and BFS strategies,  respectively. However, if the index case is exactly the ground truth, then the pair $(\textup{average error},\textup{first detection time})$ should be $(0.4,0)$ and $(0,0)$ for the DFS and BFS strategies, respectively.  If $N$ is fixed and $\mathbb{G}_N$ has a sufficiently small graph diameter, the convergence of the BFS strategy may be slower than that of the DFS strategy because the initial sets of nodes may have more neighbors. On the other hand, when $\mathbb{G}_N$ has a sufficiently large graph diameter, the DFS strategy may converge slower than the BFS strategy. Other graph-theoretic features of $\mathbb{G}_N$, such as degree and cycles, can also affect the choice of the forward contact tracing strategy. 

\subsection{Interpretation as Online Maze Solving}
We may view splitting \eqref{eq:est} into forward/backward tracing as a \emph{maze-solving} process on the epidemic network $\mathbb{G}_N$. Each step taken by the maze solver corresponds to the discovery of a leaf node $v^n$ of the rooted tree $\mathbb{G}_N$ at the $n$th iteration of the forward contact tracing stage and a corresponding $v_n^*$ of the backward contact tracing. We can treat two aforementioned sequences $S=( v_1^*,v_2^*, \ldots, v_N^*)$ and $X=( v^1,v^2, \ldots, v^N)$ as two trajectories on two mazes respectively. The trajectory $S$ is to find the desired exit $v^*_N$, and the trajectory $X$ traverses the graph. However, these two mazes are explored and solved at the same time, i.e., once the maze solver takes a step from $v^i$ to $v^{i+1}$, there is a corresponding movement on $S$ from $v_i^*$ to $v_{i+1}^*$. If we assume that there is only one contact tracer, and each time a new node is discovered in the $(i+1)$th iteration, the contact tracer has to travel from $v^{i}$ to $v^{i+1}$ with the distance $\textup{dist}(v^{i},v^{i+1})$. Then, the goal of the contact tracing is to minimize both the total traveling distance $\sum\limits_{i=1}^n \textup{dist}(v^{i},v^{i+1})$ of the contact tracer and the first detection time of $v_N^*$ (i.e., the optimal maximum-likelihood estimate had $\mathbb{G}_N$ been given in advance) during the discovery of all infected nodes in $\mathbb{G}_N$. The maze solver (i.e., contact tracer) knows neither the complete maze topology induced by $\mathbb{G}_N$ nor the target $v_N^*$. 

Upon visiting a vertex (i.e., the graph centroid of $\mathbb{G}_N$) in this abstract maze for the first time, the maze solver has information about some of its incident edges only. The contact tracer has a map of all the infected nodes visited and edges connecting them but cannot tell where each leads until $\mathbb{G}_N$ is fully traversed. Both forward and backward contact tracing mirror the process of online graph exploration or maze traversal. For a tree with no cost to discover new nodes, Theorem \ref{thm:bfs} shows that BFS finds the shortest path (i.e., sequence of centroids), while Theorem \ref{thm:DFS_main} indicates DFS offers minimal step variance. Moreover, if $\mathbb{G}_N$ is a tree, these centroids coincide with the rumor and distance centers \cite{JSTSP2018}.

Interpreting forward/backward tracing as an online maze-solving problem links (\ref{eq:est}) to classic online graph exploration \cite{onlinegraphexploration1,onlinegraphexploration2}, whose offline counterpart (Traveling Salesman Problem) is NP-hard. Thus, finding the optimal online solution of (\ref{eq:est}) is NP-hard. The best-known online algorithm (a DFS variant) achieves a constant competitive ratio of $16$ on planar graphs \cite{onlinegraphexploration1,onlinegraphexploration2}. This viewpoint suggests both low-complexity greedy approaches and more advanced algorithms for contact tracing. Additionally, machine learning can \emph{learn to optimize} such problems at scale \cite{gnnmaze1,gnnmaze2}, e.g., via GNN-based reinforcement learning. When solving (\ref{eq:est}) online with local data, the maze solver constructs a full map of graph centroids and can integrate maze-solving algorithms like A$^*$ \cite{even2011graph}. Forward tracing horizons and edge weights can be adapted, and alternate centrality measures (e.g., protection \cite{acemoglu} or vaccine centrality \cite{JSTSP2018}, Bonacich's centrality  \cite{friedkin1991centrality}) can guide novel forward-backward tracing strategies.

\section{DeepTrace: Learn to Optimize Tracing}\label{sec:gnn}

In general, solving the NP-hard problem in (\ref{eq:est1}) is computationally challenging, especially when the underlying $\mathbb{G}_N$ is large and with $N$ increasing throughout the contact tracing process. To address this challenge, we leverage GNNs to propose a scalable contact tracing algorithm named \textit{DeepTrace} (see Section \ref{sec:deep_trace}), capitalizing on the above forward and backward contact tracing decomposition. DeepTrace aims to emulate the online graph exploration strategy to maximize the number of previously unseen unique states in a limited number of steps by employing a graph-structured memory to encode the past trajectory of the contact tracer and hidden structures in the spreading model and network topologies.

\subsection{Overview of GNN-based Framework}
\label{sec:overviewGNN}
As a deep learning model for processing unstructured data, a GNN extracts graph features and information from the input data encoded as graphs. The learning mechanism of the GNN is to iteratively aggregate features and information from neighboring nodes for each node in the input graphs. Aggregating information from neighbors is equivalent to a message-passing process among nodes in a graph. Lastly, we update the values of the learning parameters for regression or classification tasks\cite{kipf2016semi, hamilton2017inductive}. To effectively capture both the statistical and graph topology attributes inherent in network-structured input data, the training stage of the GNN in semi-supervised learning is crucial. Specifically, the equations \eqref{eq:prob1} and \eqref{eq:prob2} can act as node label generators during the construction of fine-tuning datasets for the proposed algorithm.

\begin{figure*}
\centerline{\includegraphics[scale=0.28]{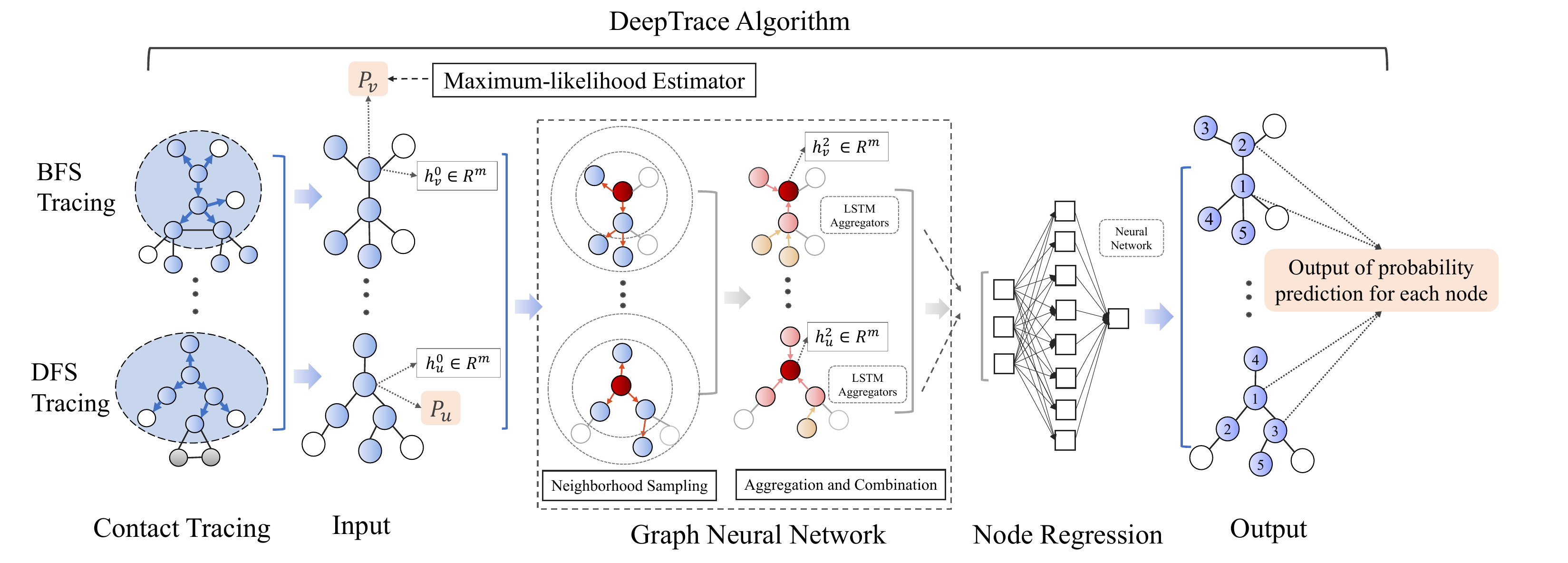}}
\captionsetup{font={footnotesize, stretch=1}, justification=raggedright}
\caption{\label{fig:gnn}The overall architecture of Algorithm DeepTrace employs a GNN, taking as input several small-scale networks obtained through BFS and DFS methods. Each node in these networks has structural features and a training label representing the permitted permutation probability. Semi-supervised learning is conducted using GraphSage with LSTM aggregators.}
\end{figure*}

To apply GNNs to solve (\ref{eq:est1}), we need to consider two key computational aspects: the underlying epidemic network $\mathbb{G}_N$ is unknown to the contact tracer, and the size of $\mathbb{G}_N$ can be potentially massive. The role of the GNN is, therefore, to facilitate the computation of a (possibly suboptimal) solution to (\ref{eq:est1}) when $\mathbb{G}_N$ is large. Due to the adaptability of GNNs, we can first generate a training set using small graphs (e.g., tens or hundreds of nodes) that are subgraphs of $\mathbb{G}_N$ and augment the descriptors with the structural features for each node of these graphs as the input data. The training stage requires that correct labels (that is, exact values of (\ref{eq:prob1}) be associated with each of these small graphs. As the size of the graph grows, the computational cost of (\ref{eq:prob1}) becomes expensive. We can instead approximate (\ref{eq:prob1}) by sampling method to efficiently compute labels for training data. Finally, we train the GNN model with the input subgraph data iteratively to update the neural network hyperparameters, as shown in Fig. \ref{fig:gnn}. We explain GNN architecture and its training in the following section.
 
\subsection{Pre-training and Fine-tuning GNNs}
\label{deeptracemethod}

The core idea for solving the backward contact tracing problem described in (\ref{eq:est1}) is to predict the value of $\sum_{\sigma \in \Omega(\mathbb{G}_{n} \mid v)} \mathbb{P}(\sigma | v)$ for each $v\in \mathbb{G}_n$ using GNNs. This prediction task can be seen as a node label prediction where the label for each node is a continuous value $\mathbb{P}(\mathbb{G}_n \mid v)=\sum_{\sigma \in \Omega(\mathbb{G}_{n} \mid v)} \mathbb{P}(\sigma | v)$. In practice, we use the log-likelihood $\log (\mathbb{P}(G_{n} | v))$ as the node label to avoid arithmetic underflow, as the values of likelihood probabilities can be extremely small as the infected network grows. We employ GraphSage \cite{hamilton2017inductive}, a widely-used inductive GNN model adaptable to varying-sized networks. We use an LSTM-based \cite{hamilton2017inductive} aggregator to gather information from the neighbors of each node, allowing the model to effectively integrate neighboring node features for learning accurate contact tracing. The $l$th layer of the GNN model is defined as follows:

  \begin{align}\label{eq:gnn}
     &\mathbf{h}^{(l)}_{N_{\mathbb{G}_N}(v)} = \mathrm{LSTM}(\{\widehat{\mathbf{w}}^{(l-1)},\mathbf{h}_u^{(l-1)}:u\in N_{\mathbb{G}_N}(v)\}),\nonumber \\
     &\mathbf{h}_v^{(l)}=\max(0,\widecheck{\mathbf{w}}^{(l)}\cdot[\mathbf{h}_v^{(l-1)};\mathbf{h}^{(l)}_{N_{\mathbb{G}_N(v)}}]),
 \end{align}
where $\widehat{\mathbf{w}}^{(l)}$ and $\widecheck{\mathbf{w}}^{(l)}$ are the learning parameters in the LSTM aggregators and combination function, respectively. The initial node feature vector is $\mathbf{h}_v^{(0)}=[\mathbf{1}, \hat{r}(v), \check{r}(v)]^T$ for $v\in \mathbb{G}_N$, which we discuss in the Section \ref{sec:feat}. To train the GNN model in DeepTrace during the two-phase training process, we define the loss functions for the pre-training and fine-tuning phases, respectively, as  follows:
\begin{align*}
  &L_{p}(\widehat{\mathbf{w}},\widecheck{\mathbf{w}},v \mid v \in  G_{n})\\
  =& \sum_{v\in G_{n}}\big|\log (\widetilde{\mathbb{P}}(G_{n} \mid v))-\mathbf{h}^{(L)}_{v}(\widehat{\mathbf{w}}, \widecheck{\mathbf{w}})\big|^2,  
\end{align*}
and
\begin{align*}
  &L_{f}(\widehat{\mathbf{w}},\widecheck{\mathbf{w}},v \mid v \in  G_{n})\\
  =& \sum_{v\in G_{n}}\big|\log (\mathbb{P}(G_{n} \mid v))-\mathbf{h}^{(L)}_{v}(\widehat{\mathbf{w}}, \widecheck{\mathbf{w}})\big|^2,  
\end{align*}
where $\widetilde{\mathbb{P}}(G_{n} \mid v)$ is the approximation of $\mathbb{P}(G_{n} \mid v)$, $\widehat{\mathbf{w}} = (\widehat{\mathbf{w}}^{(1)},\cdots,\widehat{\mathbf{w}}^{(L)})$ and $\widecheck{\mathbf{w}} = (\widecheck{\mathbf{w}}^{(1)},\cdots,\widecheck{\mathbf{w}}^{(L)})$. The effectiveness of backward contact tracing using GNN \eqref{eq:gnn} is proved as follows.

\begin{theorem}\label{thm:thm3}
 For a given epidemic network $\mathbb{G}_N=(V,E)$, denote $\textup{diam}(\mathbb{G}_N)$ as the diameter of $\mathbb{G}_N$, and define each layer of the GNN model by \eqref{eq:gnn}. Then for $\forall\; \epsilon>0$ there exist a parameter setting $(\widehat{\mathbf{w}}^*,\widecheck{\mathbf{w}}^*)$ in \eqref{eq:gnn} with at most $L = \textup{diam}(\mathbb{G}_N)+1$ layers such that
$$|\mathbf{h}^{(L)}_v-\mathbb{P}(\mathbb{G}_N \mid v)|<\epsilon, \forall\;v \in V.$$
\end{theorem}

 \textcolor{black}{
Based on the classical result on the learning capability of feedforward neural networks \cite{hornik1991approximation}, Theorem \ref{thm:thm3} states that the value of $\mathbb{P}(\mathbb{G}_{N} \mid v)$ can be approximated by GNNs with at most $\textup{diam}(\mathbb{G}_N)+1$ layers within an arbitrary error $\epsilon$. From an algorithm design perspective, Theorem \ref{thm:thm3} establishes the feasibility of approximating $\mathbb{P}(\mathbb{G}_{N} \mid v)$ using GNNs, providing the theoretical foundation for our subsequent approach to learning $\mathbb{P}(\mathbb{G}_{N} \mid v)$ with GNNs.
}

Since obtaining labeled datasets for training \eqref{eq:gnn} is computationally challenging, we employ a two-phase semi-supervised training approach to tune the learning parameters in \eqref{eq:gnn}: a pre-training phase followed by a fine-tuning phase.
\textcolor{black}{
In the pre-training phase, we train the GNN in Deeptrace using a dataset of contact tracing networks with approximations of $\mathbb{P}(G_{n} | v)$ as labels for each node. This dataset can be easily obtained,  as described in Section \ref{sec:training}. This pre-trained GNN exhibits a great enhancement in performance when compared to a GNN with randomly initialized learning parameters. Then, in the fine-tuning phase, we proceed to adjust the learning parameters in the pre-trained model using an additional dataset of contact tracing networks that contain the exact values of $\mathbb{P}(G_{n} \mid v)$ as labels for each node in the networks.  This dataset can include special cases of regularly-sized networks, such as $d$-regular graphs \cite{ShahTransIT2011, JSTSP2018}, as well as a small number of contact tracing networks that have the exact values of $\mathbb{P}(G_{n} \mid v)$ obtained by calculating probabilities of all the permitted permutations in \eqref{eq:est1} as labels for each node. Acquiring the latter dataset requires more effort and is considered highly valuable.  Therefore, it is not wasted on training the GNN from scratch but rather on fine-tuning the pre-trained model to further enhance its accuracy.  This two-phase training process enables us to effectively utilize both the dataset with approximate values of $\mathbb{P}(G_{n} \mid v)$ and the dataset with exact values of $\mathbb{P}(G_{n} \mid v)$. Thus, we can attain a GNN that exhibits enhanced performance while reducing the time and computational resources required for training.
 }

\subsection{Data Construction \& Annotation for Two-phases Training }\label{sec:training}

To construct the training datasets for pre-training and fine-tuning the GNN model, we select six different synthetic networks as the underlying network $\mathbb{G}$ with $3500$ nodes. The corresponding parameters for each type of synthetic network are detailed as follows: Erd\"{o}s R\'{e}nyi (ER) random graphs, characterized by the probability of edge formation between nodes of $10^{-3}$; Barab\'{a}si-Albert (BA) random graphs, where each new node connects to existing nodes by selecting one edge; Watts-Strogatz (WS) random graphs, in which each node connects to one nearest neighbor with a reconnection probability of $0.1$; random $3$-regular networks; stochastic block model (SBM) networks, defined by three communities, with an intra-community edge formation probability of $2^{-4}$ and an inter-community edge formation probability of $5 \times 10^{-4}$; and sensor networks \cite{SciPyProceedings_11}, which cover an area of $55 \text{km}^2$ with a communication range of $130 \text{m}$. 

Next, we apply the SI spreading model \cite{ShahTransIT2011,pd_jstsp2022} on these synthetic networks to simulate the spread of infection to generate the epidemic network $\mathbb{G}_N$. Initially, we uniformly select a source node to initiate the spread of the virus on $\mathbb{G}$. The probability of a node in the graph boundary $\mathcal{B}(G_n)$ being infected in the next time period is uniformly distributed among all outreaching edges from the infected nodes. For example, in Fig. \ref{fig:layer}, had $v_3$ been the source, the probability of $v_1$, $v_2$, $v_4$, and $v_5$ being the second infected node is $1/4$, since there are $4$ outreaching edges from $v_3$. This process continues until the number of infected nodes reaches our predetermined threshold.

\subsubsection{Dataset Construction and Annotation for Pre-training}
\label{sec:pretrain_label}
We generate $500$ underlying networks $\mathbb{G}$ with $3500$ nodes, where each $\mathbb{G}$ is uniformly selected from one of the six types of synthetic networks described above. For each $\mathbb{G}$, we construct $\mathbb{G}_N$ according to the SI model with $N$ randomly chosen from $50-1000$. For the labeled training data in the pre-training phase, we can use an approximate estimator of \eqref{eq:est} to efficiently obtain $\widetilde{\mathbb{P}}(G_{n} \mid v)$ for each node $v$ in the epidemic networks, minimizing time and effort. When the epidemic networks are degree-irregular trees, the ML estimator in \eqref{eq:est} aims to calculate the sum of all probabilities of permitted permutations. We can rewrite \eqref{eq:est} as follows:
\begin{equation}
\begin{aligned}
v\in \arg \max _{v\in G_{n}} \overline{\mathbb{P}}(G_{n} \mid v)|\Omega(G_{n} \mid v)|,
\end{aligned}
\label{eq:average}
\end{equation}
where $\overline{\mathbb{P}}(G_{n} \mid v)$ is the average of the probabilities of all permitted permutations. 
We can calculate $ |\Omega(G_{n} \mid v)| $ using a message-passing algorithm from \cite{Yu2017Boundary} with a time complexity of $\mathcal{O}(N)$. However, obtaining the exact value of $\overline{\mathbb{P}}(G_{n} \mid v)$ is challenging, as it requires identifying all the permitted permutations of $\mathbb{G}_N$ and their corresponding probabilities. Therefore, we randomly select a small sample of the permitted permutations, denoted by $\widetilde{\Omega}(G_{n} \mid v)$, and approximate $\overline{\mathbb{P}}(G_{n} \mid v)$ by averaging their probabilities. Thus, the approximate MLE problem is formulated as follows:
\begin{equation}
\begin{aligned}
\hat{v}&\in \arg \max _{v\in G_{n}} \frac{1}{|\widetilde{\Omega}(G_{n} \mid v)|}\sum_{\sigma\in \widetilde{\Omega}(G_{n} \mid v)} {\mathbb{P}}(\sigma \mid v)|\Omega(G_{n} \mid v)|\\
&=\arg \max _{v\in G_{n}} \widetilde{\mathbb{P}}(G_{n} \mid v).
\end{aligned}
\label{eq:appxp}
\end{equation}

\subsubsection{Dataset Construction and Annotation for Fine-tuning} \label{sec:fine_tune}
Since we already have a pre-trained GNN model, we can further refine it using a small amount of high-quality data. Specifically, this involves using data from epidemic networks with exact likelihood probabilities assigned to each node. We generate $250$ underlying networks $\mathbb{G}$ with $3500$ nodes, where each $\mathbb{G}$ is uniformly selected from one of the six types of synthetic networks described above. For each $\mathbb{G}$, we construct $\mathbb{G}_N$ according to the SI model with a fixed size of $N=50$. To construct a piece of labeled data with exact likelihood probability for fine-tuning, a natural idea is to calculate the likelihood probability of each node directly by using \eqref{eq:prob1} or \eqref{eq:prob2}. We first generate an epidemic network $\mathbb{G}_N$ with $N$ nodes, then for each node $v\in \mathbb{G}_N$  we generate all permitted permutation $\sigma \in \Omega(\mathbb{G}_N \mid v)$ starting from $v$ and compute $\sum_{\sigma \in \Omega(G_{n} \mid v)} \mathbb{P}(\sigma | v)$ using \eqref{eq:prob1}. Furthermore, we can construct data samples for which exact likelihood probabilities are easily obtainable using \eqref{eq:prob2}, such as the $d$-regular tree. In this case, since the probabilities of all permitted permutations are identical, it is evident from \eqref{eq:average} that the likelihood probability for a node in a $d$-regular tree, as given by \eqref{eq:est1}, is proportional to the epidemic centrality in \cite{JSTSP2018}, which can be computed using the message-passing algorithm described in \cite{JSTSP2018}.

Lastly, we repeat the data generation procedure in the pertaining phase to generate $250$ underlying networks and corresponding epidemic networks for the testing set.

\subsection{Node Features Construction}
\label{sec:feat}
\begin{figure}
\centerline{\includegraphics[scale=0.16]{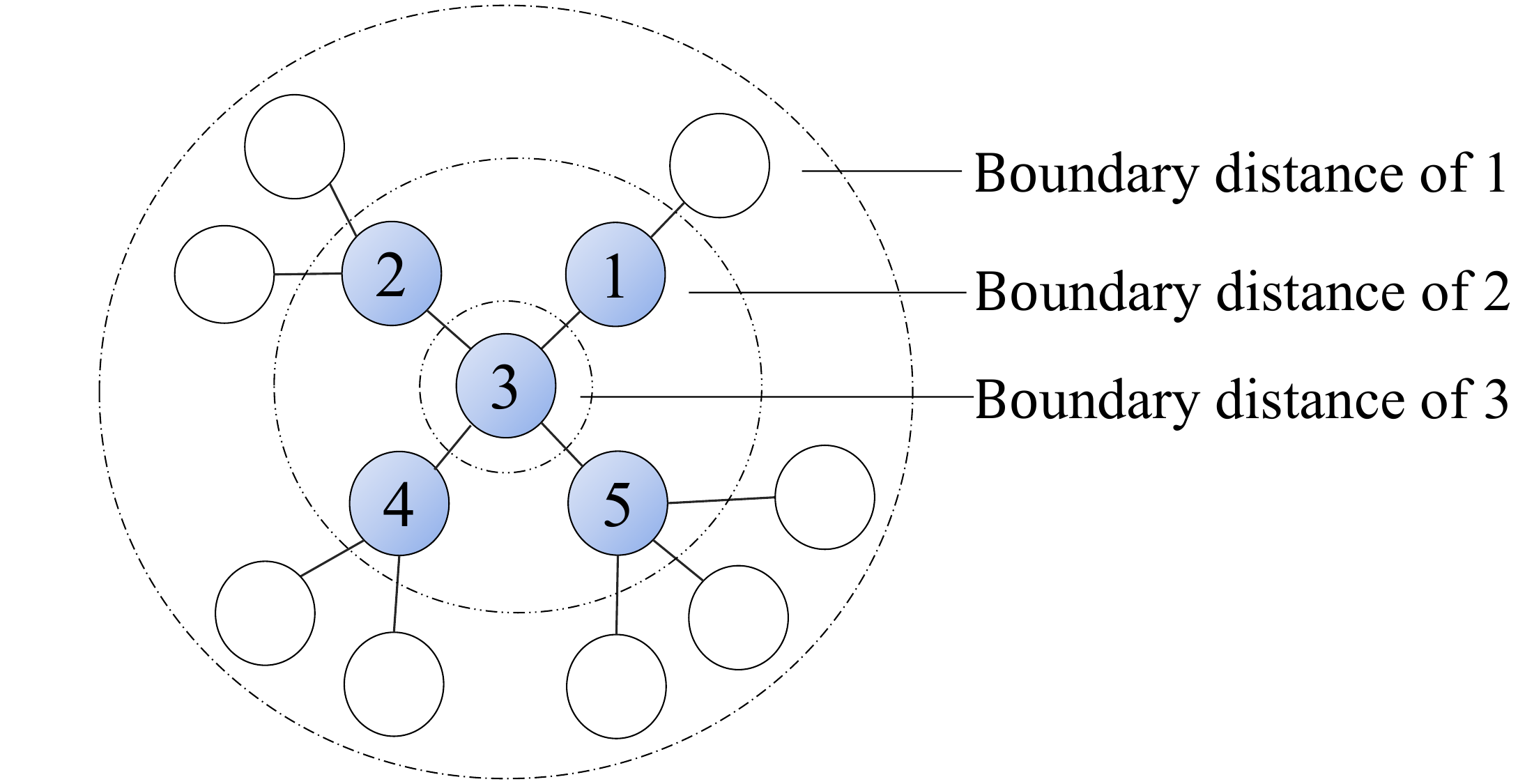}}
\captionsetup{font={footnotesize, stretch=1}, justification=raggedright}
\caption{\label{fig:layer}Illustration of the boundary distance for each node on $G_n$.}
\end{figure}

To ensure the effectiveness of GNN learning for solving (\ref{eq:est1}) in {\it Algorithm DeepTrace}, it is crucial to design the node features meticulously. This careful design enables the extraction of node information from the underlying epidemic network $\mathbb{G}_N$, corresponding to a specific spreading model during the training phase. We now consider several node features, such as the proportion of instantaneously infected nodes and boundary distances within the graph. Notably, constructing these features requires only linear computational complexity, $O(N)$, as illustrated in Fig. \ref{fig:layer}.

$\mathbf{Infected\; proportion}$: The ratio of the number of infected neighbors of a node $v_i$ to the number of all its neighbors:
$$\hat{r}(v_i) = \frac{\hat{d}(v_i)}{d(v_i)},$$
where $\hat{d}(v_i)$ is the number of infected neighbors of the node $v_i$. 
For example, the number of infected neighbors of nodes $v_1,v_2,v_3,v_4,$ and $v_5$ in Fig. \ref{fig:layer} are $1,1,4,1$ and $1$, leading to an infected proportion of nodes $v_1,v_2,v_3,v_4,$ and $v_5$ being $\frac{1}{2}, \frac{1}{3}, 1, \frac{1}{3}, $ and $\frac{1}{4}$, respectively. 
\textcolor{black}{This feature represents each node's ability to infect new individuals. Intuitively, a node with a lower infected proportion has a higher probability of infecting new individuals and a smaller probability of being the source. }Moreover, this particular node feature aims to capture the effect of the particular node position in the permitted permutation in \eqref{eq:prob1}. A node with a lower infected proportion ratio tends to be located at the front of the permitted permutation order, resulting in a smaller permutation probability. For example, as in Fig. \ref{fig:layer}, the different positions of Node $5$ for the permitted permutation $\{3,1,2,4,5\}$ and $\{5,3,1,2,4\}$ result in a probability of $1.7857\times10^{-3}$ and $1.0204\times10^{-3}$, respectively.

$\mathbf{Boundary\;distance\;ratio}$: As the infectivity of the virus evolves--newly symptomatic infectors generally exhibit higher infectivity, whereas long-term symptomatic infectors display lower infectivity--we assume that nodes at the boundary of the epidemic network are newly symptomatic infectors, while those further from the boundary are long-term symptomatic infectors. This characteristic is captured by the boundary distance ratio $\check{r}(\cdot)$. To define $\check{r}(\cdot)$, we first define the graph boundary of $G_n$ as $\mathcal{B}(G_n)=\{u\notin V(G_n) |\exists v\in G_n \text{ s.t. }  \textup{dist}(u,v)=1 \}$ and the boundary distance $b(v)$ of node $v$ on $G_n$ as $b(v)=\min\limits_{u\in \mathcal{B}(G_n)}\{\textup{dist}(u,v)\}+1$. The boundary distance ratio of $v$ on $G_n$ can now be defined as 
$$\check{r}(v) = \frac{b(v)}{\max_{v_j\in G_{n}}b(v_j)}.$$ For example, let $V(G_n)=\{v_1, v_2,v_3,v_4,v_5\}$ as shown in Fig. \ref{fig:layer}, and $\mathcal{B}(G_n)$ are those unlabeled nodes. Then, we have $b(v_1)=2$ and $b(v_3)=3$. Thus, $\check{r}(v_1)=\frac{2}{3}$, and $\check{r}(v_3)=1$. Notice that we grow $G_i$ using tree traversal algorithms; hence $G_i$ can be treated as a tree, and the boundary distance ratio of each node in $G_i$ is well-defined. From \eqref{eq:est1}, we see that finding the source also requires information on the size of the collection of all permitted permutations $\Omega(G_{n} \mid v)$. Notice that a node with a larger boundary distance ratio tends to have more permitted permutations. For example, there are $24$ permitted permutations when we choose node $v_3$ to be the source, but only $6$ permitted permutations if we choose either node $v_1, v_2, v_4$ or $v_5$ to be the source.

\subsection{The DeepTrace Algorithm}
\label{sec:deep_trace}
Inspired by the general framework discussed in Section \ref{sec:overviewGNN}, we divide the DeepTrace algorithm into two main phases: training and prediction. In the training phase, the GNN model is first pre-trained on contact tracing networks where node labels approximate the likelihood function $\widetilde{\mathbb{P}}(G_{n}|v)$. This step allows the GNN to gain an initial understanding of networked structures and learn how to approximate the infection source probabilities using these approximated labels. The second step involves fine-tuning the model using the exact values of $\mathbb{P}(G_{n}|v)$. This fine-tuning process refines the predictive capabilities of our model, enabling it to produce more accurate predictions. As the model is trained on precise source probabilities, it adjusts its parameters to optimize performance in the context of source detection. The inputs for training the GNN during the pre-training and fine-tuning phases consist of the corresponding networks that include node features. For testing the trained GNN, the inputs are the contact tracing networks from the testing dataset. The outputs from all phases--pre-training, fine-tuning, and testing are the predicted log-likelihood of each node being the source. We then find the node with the maximum output value as the predicted source.

In the prediction phase, the algorithm begins by generating the contact tracing network $G_n$ at the $n$-th stage, using either BFS or DFS. Node features $\mathbf{h}_v^0$ for each node $v\in G_n$ are initialized using $\hat{r}(v)$ and $\check{r}(v)$. These node features are then input into the fine-tuned GNN model, which predicts the likelihood of being the source for each node in $G_n$ as shown in Fig. \ref{fig:gnn}. By leveraging the trained GNN model, the algorithm efficiently computes the likelihood of each node being the source of the infection. The prediction phase allows the algorithm to trace the source in the epidemic network.

\begin{algorithm}
\SetAlgoLined
\KwIn{The index case (i.e., $G_1$); the GNN model defined in \eqref{eq:gnn} with randomly initialized parameters and contact tracing data.}
\KwOut{The log-likelihood for every node in the network to be the contagion source.}
\textbf{Training:}\\
1. Pre-train the GNN model using contact tracing networks with node labels as the approximations of $\log \widetilde{\mathbb{P}}(G_{n}|v)$ in \eqref{eq:appxp}.\\
2. Fine-tune the GNN model using contact tracing networks with node labels having the exact values of $\log{\mathbb{P}}(G_{n}|v)$ in \eqref{eq:est}.\\
\textbf{Prediction:}\\
3. \For{$n=2$ \KwTo $N$}{
Generate the contact tracing network $G_n$ using either the BFS or DFS search strategy at the $n$-th stage;\\
Construct the node features $\mathbf{h}_v^0=[\mathbf{1}, $ $\hat{r}(v), \check{r}(v)]^T$ for each node in $G_n$;\\
Predict the source probability for each node in $G_n$ using the fine-tuned GNN model in Step 2;
    }
\caption{DeepTrace}
\label{alg:calLayers}
\end{algorithm}

As contact tracing progresses, Algorithm 1 ultimately identifies the superspreaders for the entire epidemic network. The computational complexity of contact tracing in Algorithm 1 can be decomposed into three main components: sampling complexity, aggregation complexity, and the complexity of the final regression layer. Suppose there are $N$ nodes in the epidemic network. During the sampling procedure, the GNN samples a fixed-size neighborhood for each node. If each node samples $S$ neighbors and this process is repeated for $L $ layers, then the total number of nodes sampled for each node is  $O(S^L) $. 
In the final step of contact tracing, the contact tracing network contains $N$ nodes, so the sampling complexity for this network is $O(NS^L)$. Since we initiate contact tracing from the index case, the number of nodes in the contact tracing network increases sequentially. Consequently, the overall sampling complexity for the entire contact tracing process across the epidemic network is $O(N^2S^L)$.
For the aggregation procedure, the GNN aggregates information from its neighbors sampled. Assume that the feature dimension is $m$. Since the LSTM aggregator processes the features of the sampled neighbors, the complexity of the aggregation step is  $ O(mS^2)  $ per node per layer. Thus, the aggregation complexity for contact tracing the entire epidemic network is  $O(N^2LmS^2)$. In the final regression layer, a fully connected layer maps node embeddings to target values, representing the likelihood of nodes being superspreaders. The per-node complexity is $O(m)$, resulting in $O(N^2m)$ for the whole network. Thus, the overall computational complexity of contact tracing in Algorithm 1 is $O(N^2LmS^2)$.

\section{Performance Evaluation}
\label{sec:eval}
{\color{black}
In this section, we first provide simulation results on the performance of the approximate ML estimator \eqref{eq:appxp}. Secondly, we compare the BFS and DFS strategies by simulating the procedure of forward and backward contact tracing. Lastly, we demonstrate the effectiveness of {\it DeepTrace} on both synthetic networks and real-world networks based on COVID-$19$ contact tracing data in Hong Kong and Taiwan. As mentioned in Section \ref{sec:related_works}, the methods proposed in \cite{patient_zero_aaai,shah2020finding} are based on the assumptions of either SEIR or SIR spreading model, which is different from the SI model considered in this paper. We thus focus on baseline comparison relevant to the SI model, namely the rumor center heuristic in \cite{shah2012rumor}, and the SCT  (Statistical Distance-Based Contact Tracing) algorithm in \cite{pd_jstsp2022}.} 
The software for {\it Algorithm DeepTrace} and experimental data are available at \href{https://github.com/convexsoft}{https://github.com/convexsoft}.

\subsection{DeepTrace for Synthetic Networks}\label{subsec:B}
We first evaluate the performance of Algorithm DeepTrace for identifying superspreaders within epidemic networks, using the testing datasets, after we pre-train the GNN model for $150$ epochs. Then, we evaluate the performance of DeepTrace again after another $150$ epoch training for fine-tuning. There are $250$ pieces of data in the testing dataset. The generation of each piece of data follows the procedure described in Section \ref{sec:pretrain_label} except the node label in the testing dataset is the true log-likelihood $ \log{\mathbb{P}}(G_{n} | v)$.  We summarize the dataset information, including datasets for pre-training, fine-tuning, and testing, in Table \ref{tab:dataset}.

\begin{table}[htbp]\small
 \captionsetup{font={footnotesize, stretch=1}, justification=raggedright}
    \caption{Datasets of the synthetic epidemic networks for pre-training, fine-tuning, and testing when training the GNN in DeepTrace. The first column is the number of networks used in each phase.}
  \label{tab:dataset}%
\centering
\begin{tabular}{c|c|c|c|c}
\hline
No. & $|V(\mathbb{G}_N)|$               & $|E(\mathbb{G}_N)|$                        & Labels                                        & Task         \\ \hline
500       & $50\!\!\sim\!\!1000$ & $  49\!\!\sim\!\!999$ &  $ \log(\widetilde{\mathbb{P}}(G_{n} \mid v))$ & Pre-training \\ \hline 
250       & $\approx  50$        & $\approx  50$         &  $ \log({\mathbb{P}}(G_{n} \mid v))$           & Fine-tuning  \\ \hline 
250       & $50\!\!\sim\!\!1000$ & $49\!\!\sim\!\!999$   &  $ \log({\mathbb{P}}(G_{n} \mid v))$           & Testing      \\ \hline
\end{tabular}
\end{table}

\begin{figure}[htbp]
\hspace*{-0.3em}
\subfigure[]{
\begin{minipage}[t]{0.5\linewidth}

\includegraphics[width=1.8in]{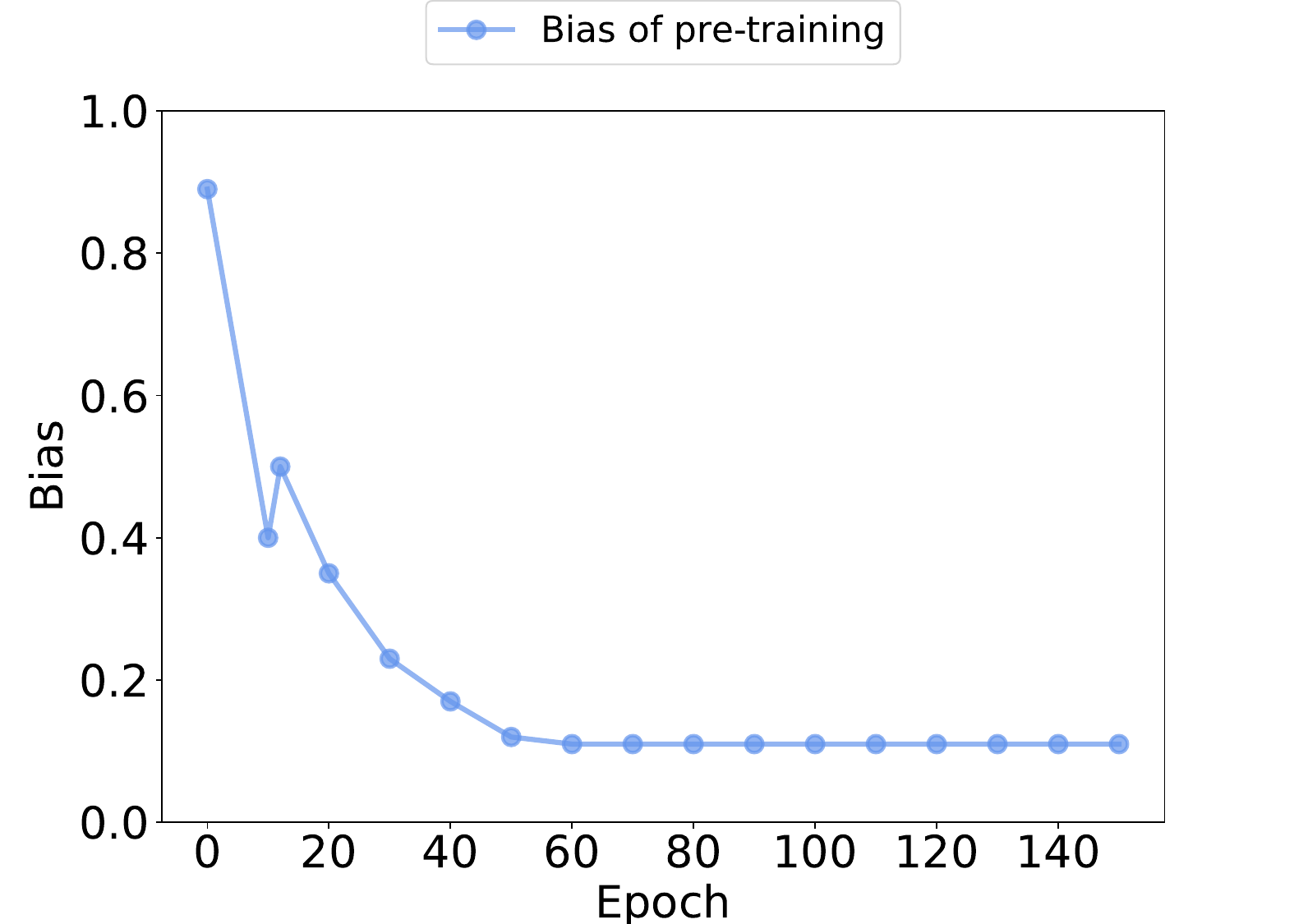}
\label{fig:bias1-a}
\end{minipage}%

\begin{minipage}[t]{0.5\linewidth}

\includegraphics[width=1.8in]{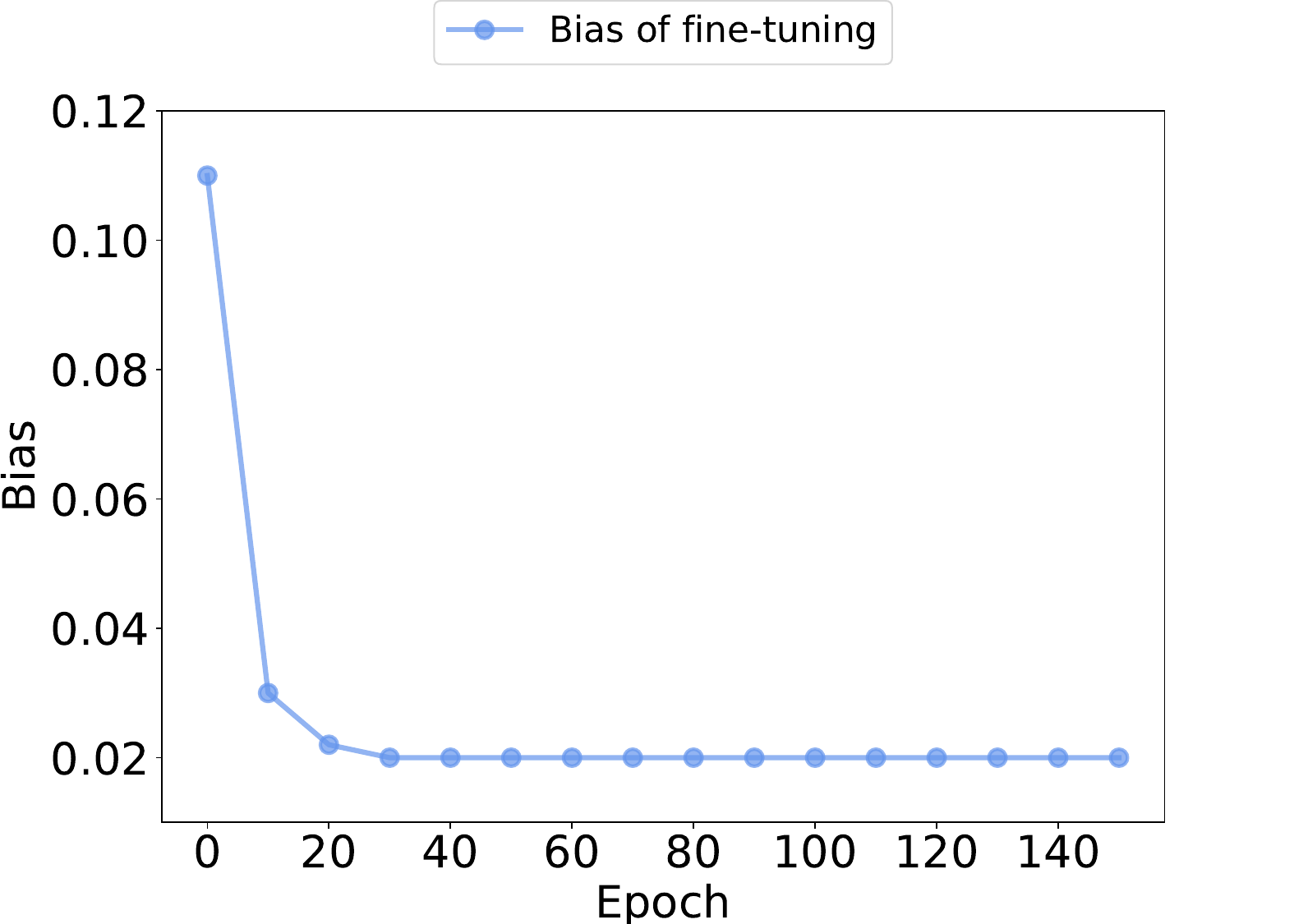}
\label{fig:bias1-b}
\end{minipage}%
}%

\hspace*{-0.3em}
\subfigure[]{
\begin{minipage}[t]{0.5\linewidth}

\includegraphics[width=1.8in]{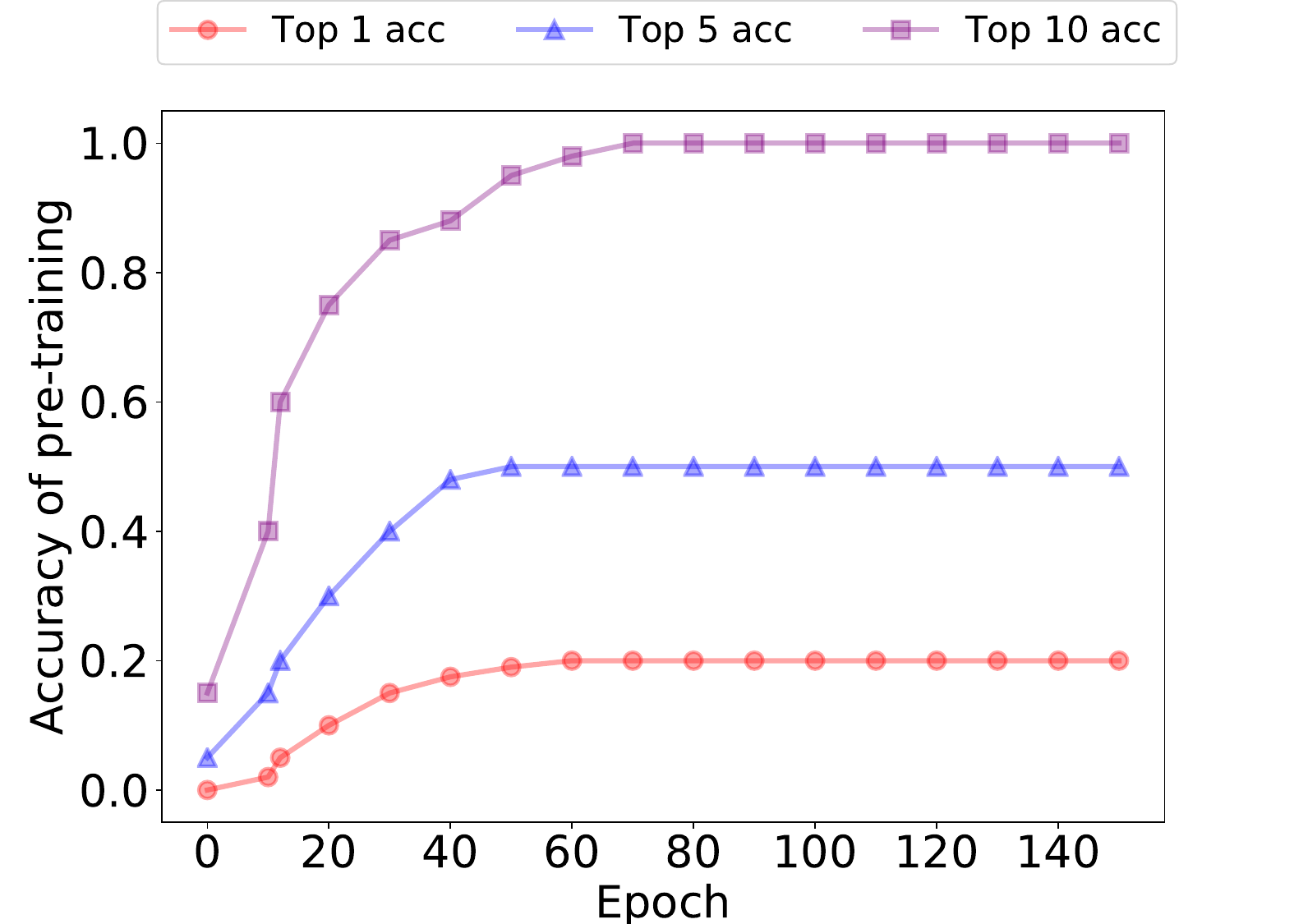}
\label{fig:bias1-c}
\end{minipage}%

\begin{minipage}[t]{0.5\linewidth}

\includegraphics[width=1.8in]{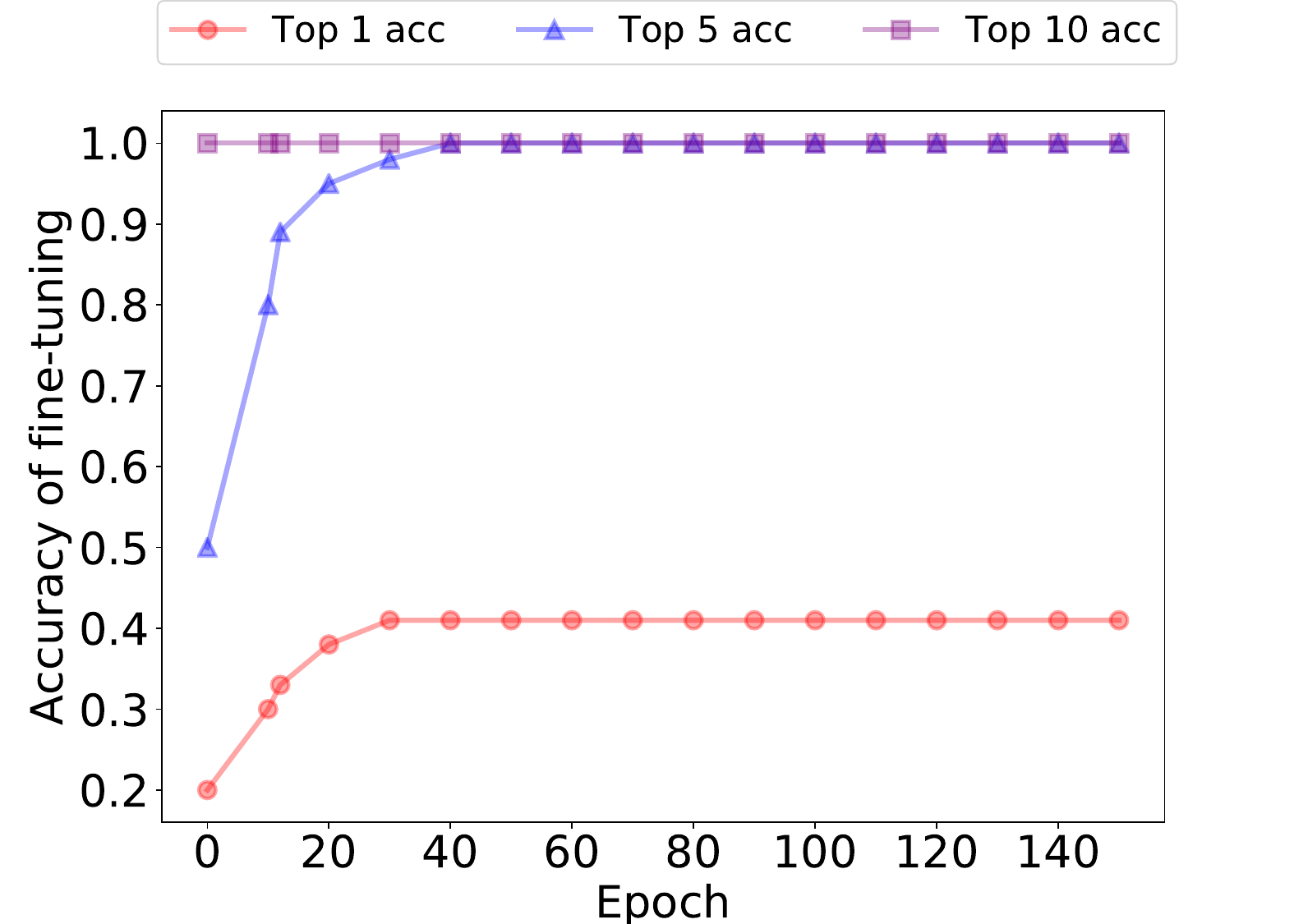}
\label{fig:bias1-d}
\end{minipage}%
}%

\centering
\captionsetup{font={footnotesize, stretch=1}, justification=raggedright}
\caption{Illustration of the results of the GNN model: (a) and (b) depict the bias trajectory between the approximate likelihood probability obtained by the GNN model and the accurate likelihood probability for the testing dataset during the pre-training and fine-tuning phases, respectively. (c) and (d) show the trajectories of top-$1$, top-$5$, and top-$10$ accuracies for the testing dataset during the pre-training and fine-tuning phases, respectively.}
\label{fig:bias1}
\end{figure}

We use the bias \cite{chen2022identifying} and a top-$k$ accuracy to evaluate the performance of the trained model in the two-phase training process. The ``top-$k$" metric signifies that the trained GNN identifies the approximate superspreader as one of the $k$ nodes with the highest probability of being superspreaders. This metric is similar to the \texttt{pass@k} metric for evaluating the functional correctness of code generated by pre-trained models in \cite{chen2021evaluating}. In particular, the \texttt{pass@k} metric is the probability that at least one of the top-$k$ generated code samples for a problem passes the unit tests \cite{chen2021evaluating}. In the context of evaluating the efficacy of pre-trained models in learning to optimize, this metric measures how well the ranked nodes in $G_n$ solve (\ref{eq:est}) based on the notion of network centrality. For example, if the approximate superspreader detected by the trained GNN is in the top-10 set, it indicates that this node is among the $10$ most likely superspreaders. Thus, the lower the value of $k$ in the top-$k$ set, the higher the likelihood that the detected node is the most probable superspreader. Moreover, a higher likelihood of the optimal solution to (\ref{eq:est}) being among the top-$k$ for a smaller $k$ implies the better performance of the pre-trained GNN. Thus, the ``top-$k$" metric effectively measures the frequency with which the global optimal solution of (\ref{eq:est}) appears within the top-ranked prediction of the pre-trained GNN, offering a quantitative assessment of the accuracy. 

Fig. \ref{fig:bias1} illustrates the value of bias and accuracy over the training epochs during both the pre-training and fine-tuning phases. Fig. \ref{fig:bias1-a} and \ref{fig:bias1-b} show that the bias decreases significantly as the number of epochs increases, with a faster and more pronounced reduction during fine-tuning. Initially, the pre-training phase exhibits higher bias, which gradually decreases and levels off after about 20 epochs, whereas fine-tuning achieves a lower bias within the first few epochs. Fig. \ref{fig:bias1-c} and \ref{fig:bias1-d} present the top-$1$, top-$5$, and top-$10$ accuracy trends, demonstrating that accuracy improves steadily throughout both phases. During pre-training, accuracy increases more slowly, with top-10 accuracy achieving the highest values, followed by top-5 and top-1 accuracy. In contrast, fine-tuning rapidly enhances accuracy, reaching a stable level within 20 epochs. These results indicate that fine-tuning not only accelerates convergence but also achieves a more optimal balance between bias reduction and accuracy improvement compared to the model with only the pre-training phase.


\begin{figure}
\centerline{\includegraphics[scale=0.24]{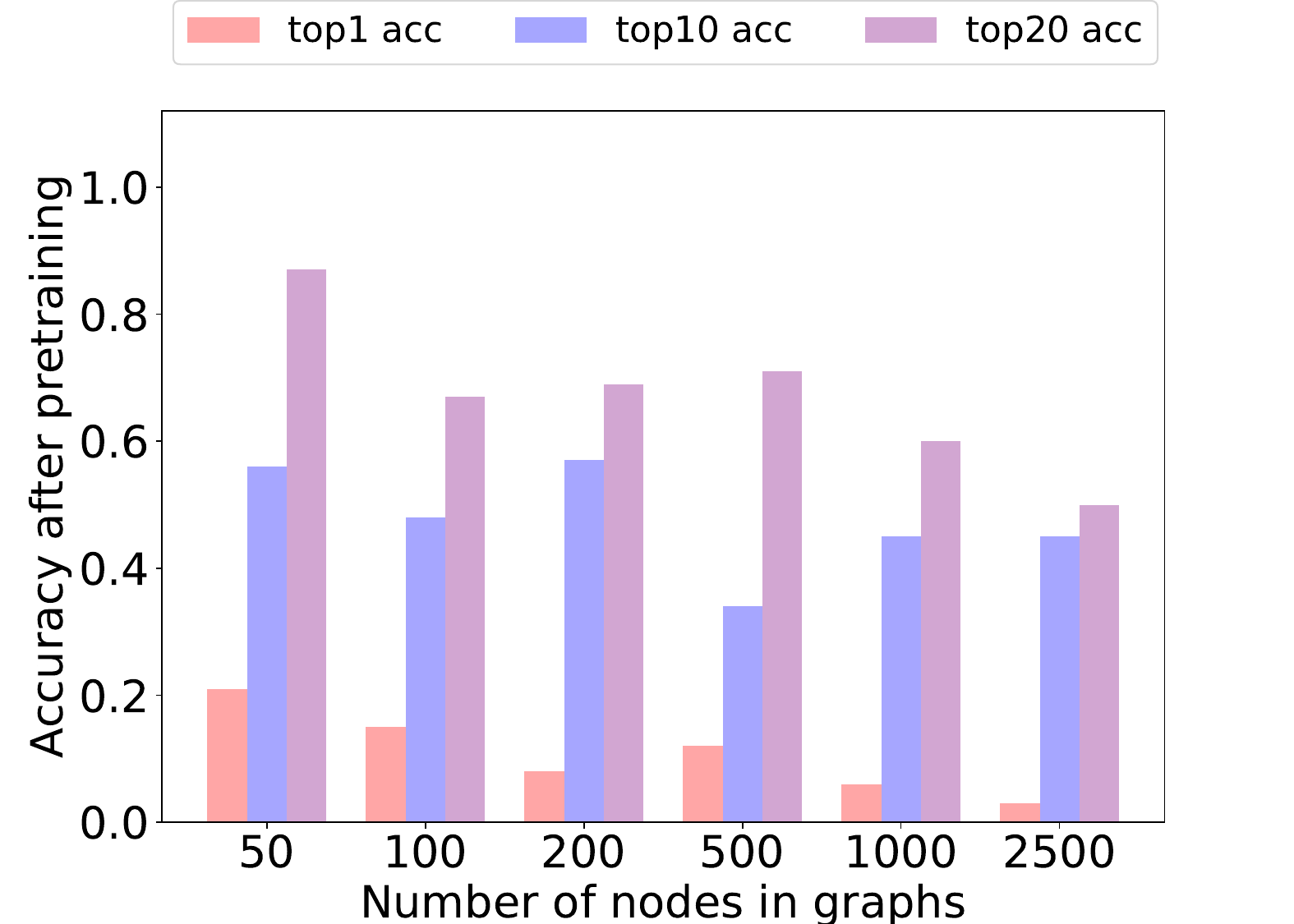}}
\captionsetup{font={footnotesize, stretch=1}, justification=raggedright}
\caption{\label{fig:pretr_gnn_acc} The top-$1$, top-$10$, and top-$20$ prediction accuracies for validation datasets after pre-training in Algorithm DeepTrace.}
\end{figure}

Another key evaluation is to measure the generalizability of {\it DeepTrace Algorithm}, that is, to validate the effectiveness of the trained model by the small-scale epidemic networks in the larger epidemic networks. We construct six additional synthetic network validation datasets, each with $100$ random $3$-regular networks, and the number of nodes in each set of networks is $50, 100, 200, 500, 1000,$ and $2500$, respectively. We use the SI model to simulate the spreading for a given number of infected nodes to generate the infection graphs (see Section \ref{sec:training}). We evaluate the trained GNN's performance on validation datasets using top-$k$ accuracy.

\begin{figure}
\centerline{\includegraphics[scale=0.24]{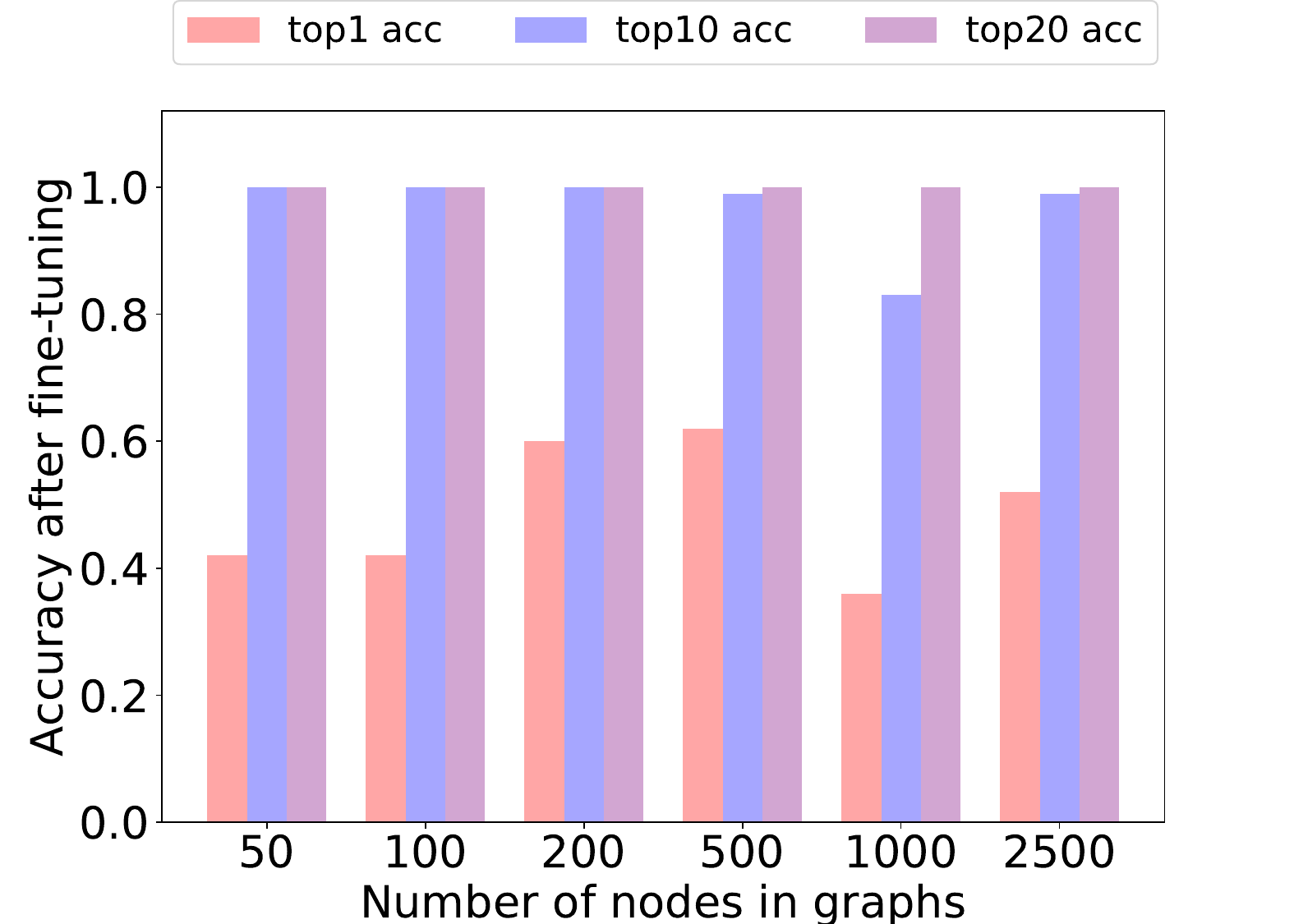}}
\captionsetup{font={footnotesize, stretch=1}, justification=raggedright}
\caption{\label{fig:exp_gnn_acc} Validation dataset accuracy metrics at top-$1$, top-$10$, and top-$20$ prediction levels after fine-tuning in Algorithm DeepTrace.}
\end{figure}

After training for 150 epochs, we evaluate the prediction accuracy at top-1, top-10, and top-20 for each validation dataset. The results of the pre-training and fine-tuning processes are illustrated in Fig. \ref{fig:pretr_gnn_acc} and Fig. \ref{fig:exp_gnn_acc}, respectively. Fig. \ref{fig:pretr_gnn_acc} displays the accuracy trends during the pre-training phase, showing a steady improvement in performance as the model learns. In contrast, Fig. \ref{fig:exp_gnn_acc} depicts the fine-tuning phase results, highlighting how the model adjusts and refines its predictions with additional training. These figures demonstrate that the training accuracy remains relatively stable, not deteriorating significantly even as the number of nodes in the epidemic networks increases. This demonstrates the GNN's robustness and adaptability across network sizes, highlighting the utility of training on small networks for generalization to larger ones. Training on smaller networks reduces time and enables transfer learning for source detection in larger networks.

\subsection{Evaluation on BFS and DFS tracing}
{\color{black}
In this section, we compare the performance of BFS and DFS tracing strategies using first detection time and average error as evaluation metrics, defined by equations \eqref{eq:avg_err} and \eqref{eq:first_dect}, respectively. 
Here, we consider five groups of epidemic network datasets with distinct topological structures. The relevant statistics for these data are displayed in TABLE \ref{tab:dataset2}. In the final column, we use $d$ to represent the degree of the regular network, $p$ to indicate the likelihood of edge formation between nodes in the ER random network, and $NC$/$p$/$q$ to denote the number of communities, the probability of edge formation within a community, and the probability of edge formation between communities, respectively. Additionally, we define the sensor network's coverage area as $Area$ and the communication range, denoted as $CR$, as the utmost distance between sensors that can form an edge.
Let us describe the simulation experiment process. In synthetic networks, we first establish the underlying network $\mathbb{G}$ using different random network models. Next, based on the SI model described in Section \ref{sec:SI_Model}, we uniformly and randomly select a source node and begin simulating the spread of the virus on $\mathbb{G}$. Since the number of nodes in our real-world network data is around 100, we stop this spreading process once the number of infected nodes reaches 100, resulting in a 100-node epidemic network $\mathbb{G}_N$. To ensure fairness (as different index nodes may lead to significant performance differences), each node in $\mathbb{G}_N$ is selected as the index node to start the contact tracing process.

We perform forward contact tracing using BFS or DFS and employ Algorithm DeepTrace along with two other methods, Rumor Center \cite{shah2012rumor}, and SCT \cite{pd_jstsp2022}, for backward contact tracing. As we expand the contact tracing network $G_n$, each time a new node is added, we compute the current estimated superspreader in $G_n$ and measure the distance between this estimated superspreader and the true source, noting the error in hops at that moment (cf. \eqref{eq:avg_err}). When the estimated superspreader first corresponds to the true source, we record the current size of $G_n$ as the first detection time. When $G_n = \mathbb{G}_N$, we will have identified $100$ approximated superspreaders and their corresponding error hops. We calculate the average of these 100 error hops as the average error hops for this simulation experiment. This experiment is repeated $100$  times on the same underlying network $\mathbb{G}$, but with different epidemic networks $\mathbb{G}_N$, recording the average error hops, first detection time, and the time spent. We initiate the contact tracing simulation directly on this data for the real-world network data (Taiwan and Hong Kong COVID-19 datasets), as these networks already represent epidemic conditions.

The theoretical result we obtained in Theorem \ref{thm:bfs} for the BFS strategy describe the distance between the approximated superspreader and the MLE, whereas the theoretical results from Lemma \ref{lem:comp_S} and Theorem \ref{thm:DFS_main}, for the DFS strategy calculate the upper bound of the number of nodes that will be marked as approximated superspreaders during the contact tracing process. The theoretical results for BFS and DFS are entirely different in scope, so we cannot determine which strategy is better based on theory alone. In our simulation experiments, we tested six combinations of DeepTrace, Rumor Center, and STC with either BFS or DFS and recorded the first detection time and the average distance error in Table \ref{tab:combined}. We can observe that DeepTrace+BFS outperforms other methods from the perspective of the first detection time. Additionally, within Rumor Center and STC, the BFS strategy generally outperforms the DFS strategy. On the other hand, from the perspective of the average distance error, when the underlying network is a tree with a symmetrical structure, such as complete N-ary graphs or regular trees, DeepTrace+DFS performs better. In summary, DeepTrace generally exhibits lower average distance errors than the baseline methods. BFS tracing often performs better than DFS for DeepTrace, indicating that BFS may be more effective for these structures. The results show that DeepTrace achieves higher accuracy in complex network topologies and consistently outperforms baseline methods across various networks.} We evaluated the computational efficiency of DeepTrace, Rumor Center, and STC, measuring the average contact tracing time using DFS and BFS. As shown in Table \ref{tab:efficiency_comp}, DeepTrace achieves significantly higher computational speed than the other methods.

\begin{table*}\color{black}\small
\captionsetup{font={footnotesize, stretch=1}, justification=raggedright}
\caption{Summary of the datasets used for performance comparison with BFS and DFS tracing.}
\label{tab:dataset2}
\centering
\begin{tabular}{c|c|c|c|p{2.51cm}<{\centering}}
\hline
Epidemic network             & Underlying graph                                                     & $|\mathbb{G}_N|$ & Components & Other parameters        \\ \hline
Complete N-ary graph         & \begin{tabular}[c]{@{}c@{}}Nodes = 3500, Edges = 3499\end{tabular} & 100      & 1          & $N=3$         \\ \hline
Regular tree                 & \begin{tabular}[c]{@{}c@{}}Nodes = 3500, Edges = 3499\end{tabular} & 100       & 1          & $d = 3$       \\ \hline
ER random graph              & \begin{tabular}[c]{@{}c@{}}Nodes = 3500, Edges = 6016\end{tabular} & 100      & 1          & $p = 10^{-3}$ \\ \hline
SBM network              & \begin{tabular}[c]{@{}c@{}}Nodes = 3500, Edges = 8086\end{tabular} & 100      & 1          & $p=5\!\!\times\!\! 10^{-4}$, $q= 2\!\!\times\!\! 10^{-4}$, NC=$3$ \\ \hline
Sensor network             & \begin{tabular}[c]{@{}c@{}}Nodes = 3500, Edges = 5710\end{tabular} & 100       & 1          & Area=$55km^2$, CR=$130m$  \\ \hline
Real-world network (Taiwan) & \begin{tabular}[c]{@{}c@{}}Nodes =92, Edges =117\end{tabular}       & 92        & 1          & --            \\ \hline
\end{tabular}
\end{table*}

\begin{table*}\color{black}\small
\captionsetup{font={footnotesize, stretch=1}, justification=raggedright}
\caption{Comparison of Average First Detection Time and Average Distance Error for DeepTrace and baseline methods across different epidemic networks, using BFS and DFS tracing strategies.}
\label{tab:combined}
\centering
\begin{tabular}{c|cc|cc|cc||cc|cc|cc}
\hline
\multirow{3}{*}{Epidemic Network} & \multicolumn{6}{c||}{Average First Detection Time}                 & \multicolumn{6}{c}{Average Distance Error}                          \\ \cline{2-13} 
                                  & \multicolumn{2}{c|}{DeepTrace} & \multicolumn{2}{c|}{Rumor Center} & \multicolumn{2}{c||}{SCT} & \multicolumn{2}{c|}{DeepTrace} & \multicolumn{2}{c|}{Rumor Center} & \multicolumn{2}{c}{SCT} \\ \cline{2-13} 
                                  & \multicolumn{1}{c|}{DFS}   & BFS   & \multicolumn{1}{c|}{DFS}   & BFS   & \multicolumn{1}{c|}{DFS} & BFS   & \multicolumn{1}{c|}{DFS}  & BFS  & \multicolumn{1}{c|}{DFS}  & BFS  & \multicolumn{1}{c|}{DFS}  & BFS  \\ \hline
Complete N-ary graph              & \multicolumn{1}{c|}{34.00} & \textbf{7.17}  & \multicolumn{1}{c|}{34.81} & 8.18  & \multicolumn{1}{c|}{38.60}    & 16.95  & \multicolumn{1}{c|}{\textbf{0.84}} & 1.54  & \multicolumn{1}{c|}{1.35} & 2.31  & \multicolumn{1}{c|}{1.04} & 1.66  \\ \hline
Regular tree                      & \multicolumn{1}{c|}{36.10} & \textbf{16.14}  & \multicolumn{1}{c|}{63.06} & 18.21 & \multicolumn{1}{c|}{64.44}    & 46.25  & \multicolumn{1}{c|}{\textbf{2.37}} & 2.45  & \multicolumn{1}{c|}{2.63} & 2.56  & \multicolumn{1}{c|}{2.62} & 3.30  \\ \hline
ER random graph                   & \multicolumn{1}{c|}{48.27} & 12.48 & \multicolumn{1}{c|}{65.13} & \textbf{10.97} & \multicolumn{1}{c|}{23.56}    & 36.75  & \multicolumn{1}{c|}{2.70} & \textbf{2.52}  & \multicolumn{1}{c|}{2.93} & 2.92  & \multicolumn{1}{c|}{2.78} & 2.55  \\ \hline
SBM network                       & \multicolumn{1}{c|}{\textbf{34.07}} & 47.14 & \multicolumn{1}{c|}{49.03} & 54.98 & \multicolumn{1}{c|}{44.56}    & 40.70  & \multicolumn{1}{c|}{2.55} & \textbf{2.30}  & \multicolumn{1}{c|}{2.78} & 2.86  & \multicolumn{1}{c|}{2.73} & 2.69  \\ \hline
Sensor network                    & \multicolumn{1}{c|}{32.47} & \textbf{18.71} & \multicolumn{1}{c|}{33.45} & 21.14 & \multicolumn{1}{c|}{39.28}    & 36.75  & \multicolumn{1}{c|}{2.67} & \textbf{2.61}  & \multicolumn{1}{c|}{3.50} & 3.42  & \multicolumn{1}{c|}{3.06} & 2.85  \\ \hline
Real-world network (Taiwan)       & \multicolumn{1}{c|}{1.18}  & \textbf{0.55}  & \multicolumn{1}{c|}{2.95}  & 3.08  & \multicolumn{1}{c|}{4.59}    & 4.61   & \multicolumn{1}{c|}{0.97} & \textbf{0.65}  & \multicolumn{1}{c|}{1.38} & 0.88  & \multicolumn{1}{c|}{0.79} & 0.80  \\ \hline
\end{tabular}
\end{table*}

\begin{table}\color{black}\small
\captionsetup{font={footnotesize, stretch=1}, justification=raggedright}
\caption{The efficiency comparison, in terms of time required (measured in seconds) for contact tracing, between Algorithm DeepTrace and other baseline methods across epidemic networks of varying scales.}
\label{tab:efficiency_comp}
\centering
\begin{tabular}{c|p{1.21cm}<{\centering}|p{1.21cm}<{\centering}|p{1.21cm}<{\centering}|p{1.21cm}<{\centering}}
\hline
  Methods           & N = 64 & N = 128 & N=256 & N=512 \\ \hline
DeepTrace    &    $\boldsymbol{0.58}$    &   $\boldsymbol{5.83\!\!\times\!\!10^1}$       &    $\boldsymbol{7.39\!\!\times\!\!10^2}$    &   $\boldsymbol{1.12\!\!\times\!\!10^3}$    \\ \hline
Rumor center & $2.32\!\!\times\!\! 10^1 $  &   $4.36\!\!\times\!\! 10^2 $     &  $1.03\!\!\times\!\!10^3$     &  $2.52\!\times \!10^4$     \\ \hline
SCT          & $4.32\!\times \!10^1 $  &   $6.55\!\!\times\!\! 10^2 $    &   $6.22\!\times \!10^3$    &    $1.43\!\times \!10^5$     \\ \hline
\end{tabular}

\end{table}


\subsection{DeepTrace for COVID-$19$ Epidemic Networks}
In this section, we conduct experiments on COVID-$19$ pandemic data in Taiwan and Hong Kong to evaluate the performance of {\it Algorithm DeepTrace}. A summary of these epidemic network datasets is presented in TABLE \ref{tab:dataset3}. We refer to $G_n$ as the epidemic network in the following, where nodes in $G_n$ represent confirmed cases, and each edge in $G_n$ implies that there is close contact between two confirmed cases. We compare {\it Algorithm DeepTrace} to estimators using \eqref{eq:est1}, and since these two clusters are relatively small, the computation of the ML estimator \eqref{eq:est1} is feasible.


\begin{table*}[htbp]\color{black} \small
\captionsetup{font={footnotesize, stretch=1}, justification=raggedright}
    \caption{A summary of COVID-$19$ epidemic network datasets for Algorithm DeepTrace.}
  \label{tab:dataset3}%
  \centering
\begin{tabular}{p{3.25cm}<{\centering}|p{0.95cm}<{\centering}| p{0.95cm}<{\centering} |p{1.45cm}<{\centering}| p{1.85cm}<{\centering} | p{1.85cm}<{\centering} |p{2.05cm}<{\centering}}
\hline
Epidemic network                  & Nodes & Edges & Components & Ground truths & Avg. Error Hops to MLE & Avg. Err. Hops to ground truth \\ \hline
Taiwan  (2022.03.13-2022.04.01)   & 92   & 117   & 1          & Known         & 0                      & 0                              \\ \hline
Hong Kong (2021.12.31-2022.01.22) & 102   & 97    & 4          & Known         & 0                      & 1/4                            \\ \hline
Hong Kong (2022.01.31-2022.02.03) & 402   & 390   & 11         & Unknown       & 0                      & ground truth unknown                             \\ \hline
\end{tabular}

\end{table*}%

\subsubsection{Contact Tracing for COVID-$19$ Pandemic in Taiwan}

The COVID-19 Omicron is a variant of the SARS-CoV-2 virus first reported on 24 November 2021, has shorter incubation times and is known to be more contagious than previous variants (with an effective reproduction number of $18.6$ as compared to the Alpha variant of $3$) that has since become the predominant variant in circulation worldwide. The infectiousness of Omicron has led to huge infection graphs and severely undermined the capacity of contact tracing in many countries. We first conduct experiments using data on the early spread of Omicron of COVID-$19$ in Taiwan from March 13, 2022, to April 1, 2022. The data is collected from the confirmed case reports released daily by the Taiwan Centers for Disease Control at \href{https://www.cdc.gov.tw}{https://www.cdc.gov.tw}. 
Each node in the networks represents a confirmed case and is marked with the reported number of the confirmed case. Each edge represents a close contact between two confirmed cases.

\begin{figure}[h]
\centerline{\includegraphics[scale=0.28]{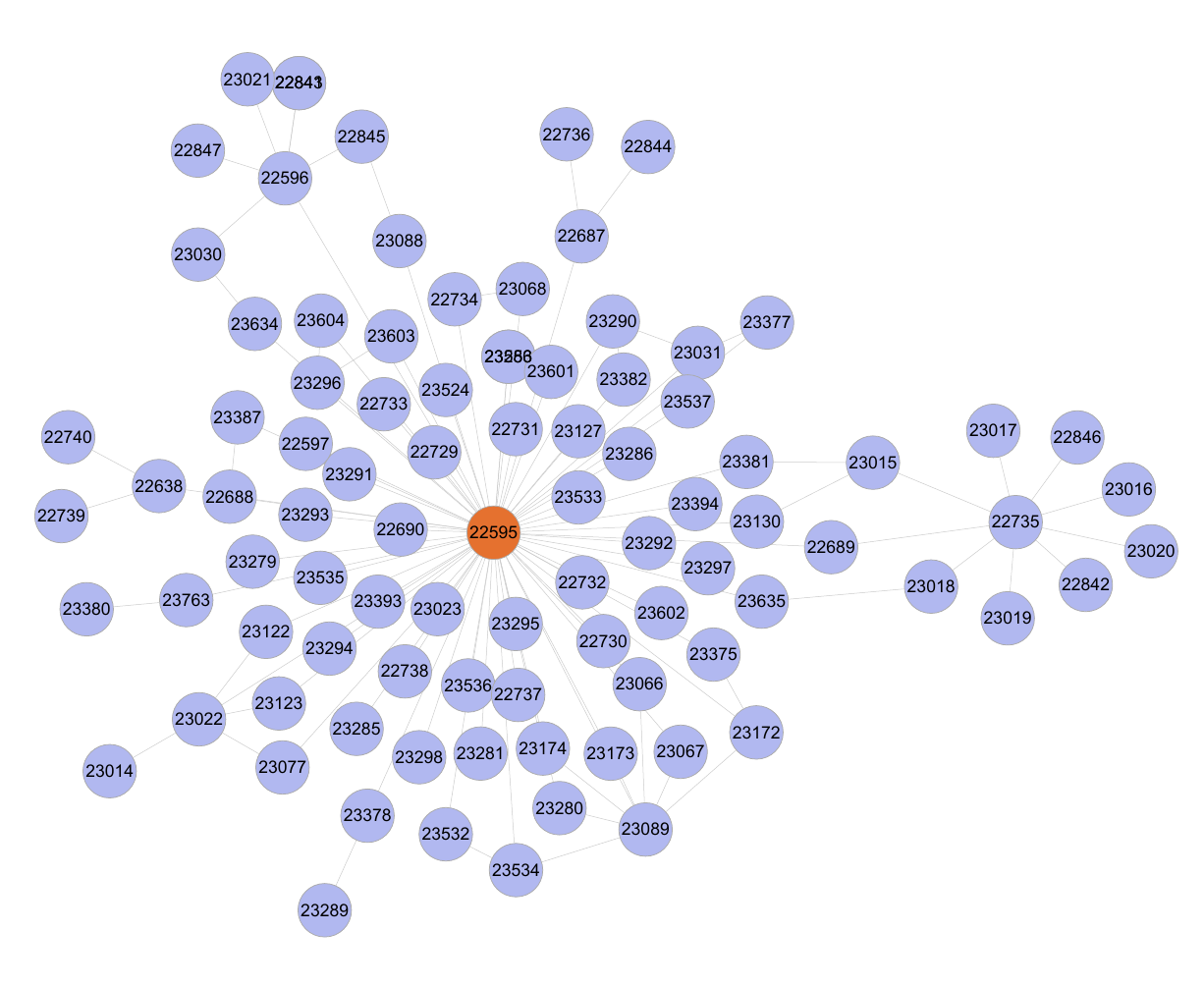}}
\captionsetup{font={footnotesize, stretch=1}, justification=raggedright}
\caption{\label{fig:Omicron_tw} 
Illustration of contact tracing with DeepTrace for a COVID-19 Omicron variant outbreak in Taiwan in early 2022. The blue nodes represent infected nodes, while the red nodes are the superspreaders predicted by DeepTrace, which incidentally correspond exactly to the actual superspreaders (i.e., ground truth) reported by the Taiwanese authorities.}
\end{figure}

\textcolor{black}{Fig. \ref{fig:Omicron_tw} is the largest contact tracing network constructed with the collected data with $92$ confirmed cases. Case 22595 is the reported superspreader in this contact tracing network. As Case 22595 is also the earliest discovered case, we consider it the index case for contact tracing. Using the trained GNN in Algorithm DeepTrace, we applied it to this network and detected that Case 22595, colored red, is the most likely superspreader. These findings align with the actual superspreaders reported by the Taiwanese authorities.}

\subsubsection{Contact Tracing for COVID-$19$ Pandemic in Hong Kong}



Due to Hong Kong's strict ``Zero-COVID" policy with rigorous quarantine and testing, its population remained largely unaffected by COVID-19 until January 2022. However, the arrival of the Omicron variant led to one of the world's highest death tolls in early 2022, despite widespread vaccine access and Omicron's lower lethality \cite{xie2023resurgence,mefsin2022epidemiology}. This unique scenario allows for real-world application of the SI spreading model and evaluation of Algorithm DeepTrace using data from Hong Kong's Omicron outbreak. We first consider an epidemic network with four connected components in Hong Kong from 31 December 2021 to 22 January, the data for which is from \cite{mefsin2022epidemiology}. This epidemic network contains 102 SARS-CoV-2 Omicron infections associated with four imported cases (nodes 1, 29, 70, and 100) in the connected components, which serve as ground truths of superspreaders. 

\begin{figure}[h]
\centerline{\includegraphics[scale=0.23]{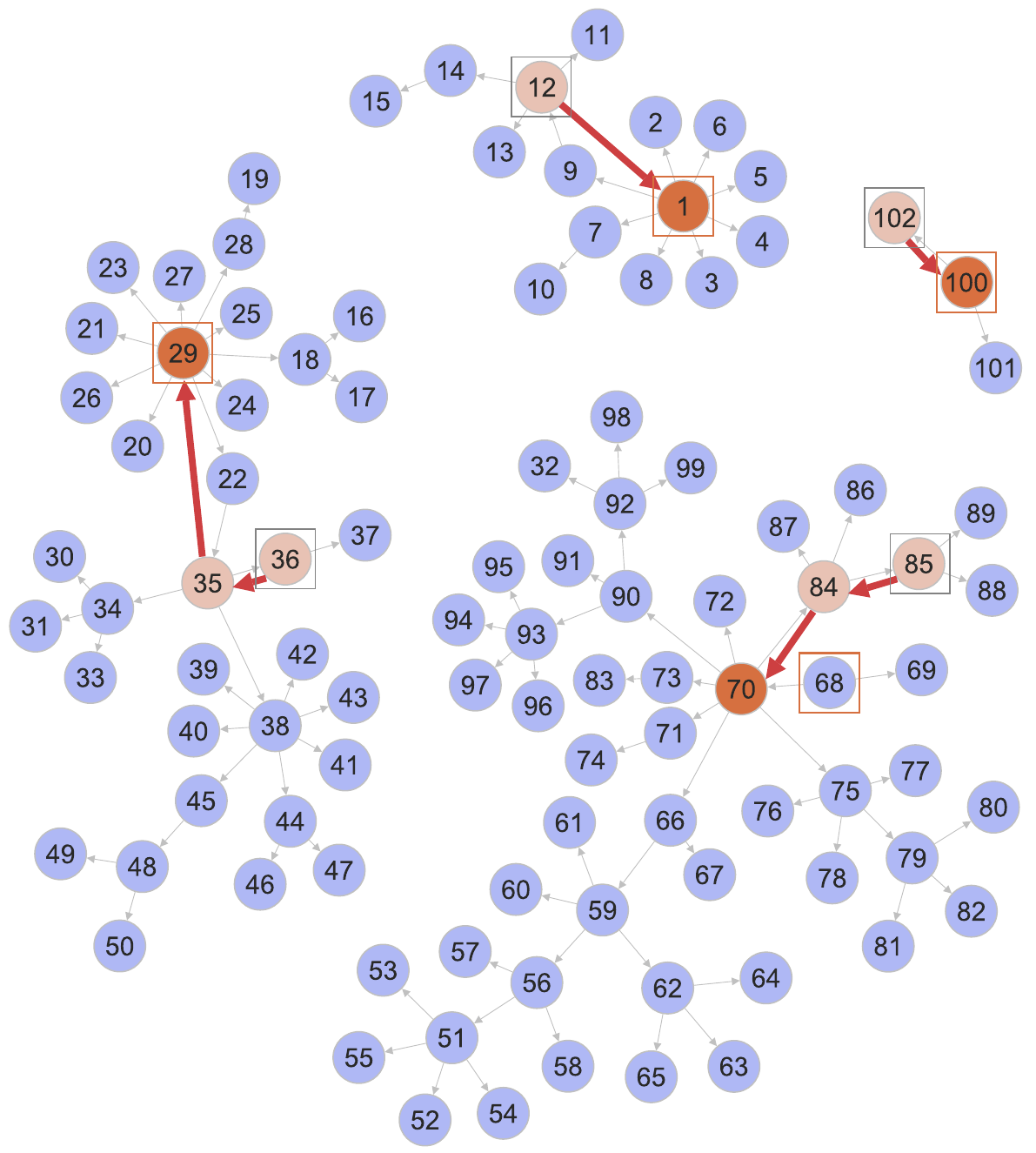}}
\captionsetup{font={footnotesize, stretch=1}, justification=raggedright}
\caption{\label{fig:hk_source} 
\textcolor{black}{Illustration of contact tracing with DeepTrace for an early Omicron outbreak in Hong Kong from 31 December 2021 to 22 January. The blue nodes represent infected nodes. The nodes in the gray boxes are considered the index cases of the Omicron outbreak. The nodes in the red boxes are the true superspreaders. The nodes in pink are superspreaders detected in the subgraph of the epidemic network during contact tracing.  The nodes in red are the most likely superspreaders in the spreading clusters in the epidemic network. The red arrows indicate the movement of superspreaders detected by DeepTrace as contact tracing continues.}}
\end{figure}

DeepTrace is used to iteratively identify the most likely superspreader, as outlined in \eqref{eq:est1}, while the forward contact tracing network expands. Fig. \ref{fig:hk_source} illustrates the epidemic network and contact tracing process. Initially, nodes $12$, $36$, $85$, and $102$ serve as index cases for the tracing process. As the network grows, DeepTrace identifies the most likely superspreaders, shifting from node $12$ to $1$, from $36$ to $35$, from $85$ to $84$ and then $70$, and from $102$ to $100$ across the four connected components. The final detected superspreaders (nodes $1$, $29$, $70$, and $100$) correspond to the predictions from \eqref{eq:est1}. Nodes $1$, $29$, and $100$ are confirmed superspreaders, while node $70$ is one hop away from the true superspreader, node $68$. These results corroborate the efficiency of DeepTrace.

Next, we consider utilizing DeepTrace on another epidemic network in Hong Kong, where the ground truths of superspreaders are unknown. Following \cite{xie2023resurgence}, we retrieved data of confirmed Omicron cases in Hong Kong from January 31, 2022, to February 3, 2022, from the Hong Kong government's public sector open data portal at \href{https://data.gov.hk/en-data/dataset/hk-dh-chpsebcddr-novel-infectious-agent}{https://data.gov.hk/en-data/dataset/hk-dh-chpsebcddr-novel-infectious-agent}. We constructed epidemic networks using this data, where each node represents a confirmed case and edges connect nodes that visited the same building. The earliest reported case in \cite{xie2023resurgence} served as the initial index case for forward contact tracing, with DeepTrace iteratively identifying superspreaders as the network expanded.

\begin{figure}[h]
\centerline{\includegraphics[scale=0.35]{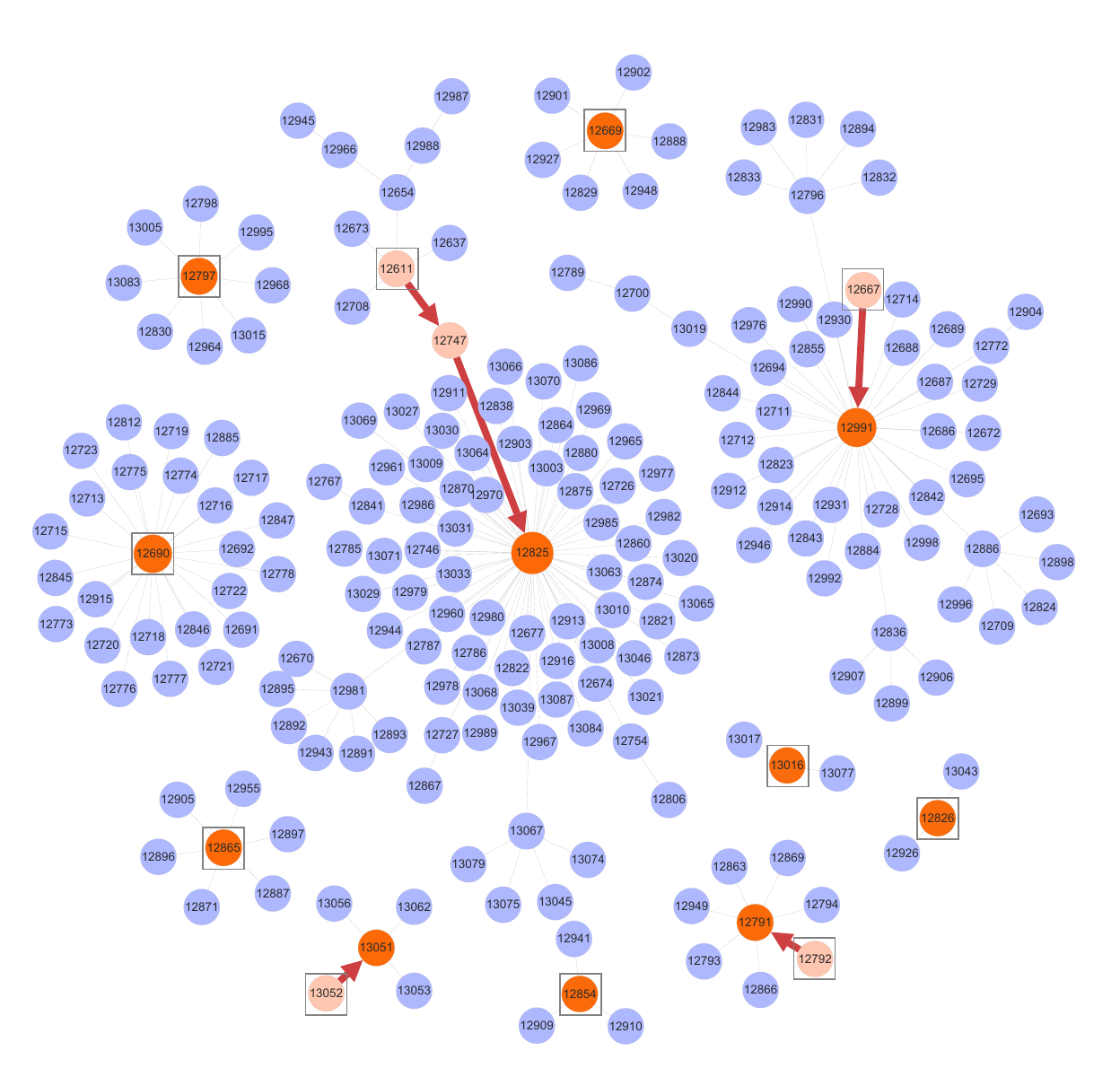}}
\captionsetup{font={footnotesize, stretch=1}, justification=raggedright}
\caption{\label{fig:Omicron_hk} 
Illustration of contact tracing using {\it Algorithm DeepTrace} for an early Omicron outbreak in Hong Kong in 2022. The blue nodes represent infected nodes. The nodes in the boxes are the index cases of the Omicron outbreak. The nodes in pink are superspreaders detected in the subgraph of the epidemic network during contact tracing.  The nodes in red are the most likely superspreaders in the spreading clusters in the epidemic network. The red arrows indicate the movement of superspreaders detected by DeepTrace as contact tracing continues.}
\end{figure}

Fig. \ref{fig:Omicron_hk} illustrates the contact tracing networks built from the retrieved data, showing how DeepTrace's iterative outputs converge on the most likely superspreader from \eqref{eq:est1} as the network expands. For instance, in the largest spreading cluster, case 12611 is the index case and initial superspreader. As the tracing network develops, DeepTrace identifies the superspreader, shifting from case 12611 to 12747 and finally to case 12825, which matches the prediction from \eqref{eq:est1}. This demonstrates that DeepTrace effectively identifies likely superspreaders without prior knowledge of the true ones. Targeting these identified superspreaders with treatment or quarantine measures could help suppress further epidemic spread.

\section{Conclusion}\label{sec:conclusion}
Digital contact tracing remains a crucial challenge in the pandemic response. This paper formulates both forward and backward contact tracing strategies as maximum likelihood estimations of the epidemic source, which are essential for identifying superspreaders. Starting from an index case, contact tracers explore the epidemic network using breadth-first or depth-first search to compute the graph center of the contact tracing subgraph, leading to optimal likelihood estimates for specific topologies. We introduce Algorithm DeepTrace, which employs a graph neural network with semi-supervised learning, incorporating pre-training and fine-tuning phases for large graphs through message passing. This approach enhances large-scale graph inference in computational epidemiology and connects to online graph exploration, providing low-complexity greedy algorithms for contact tracing. We validate our methodology by modeling real COVID-19 superspreading events in Hong Kong and Taiwan, showing that our algorithm outperforms existing heuristics.

\textcolor{black}{
To enhance digital contact tracing and address challenges like Antimicrobial Resistance (AMR), future work will focus on advanced graph exploration and adaptive learning to handle new epidemic and AMR networks. We aim to develop efficient decomposition techniques for optimization, extend Algorithm DeepTrace to process real-time data streams, and improve scalability for larger networks. By integrating pathogen resistance data, our approach can optimize contact tracing and contribute to scalable, data-driven surveillance systems for pandemics and AMR mitigation.}

\section{Acknowledgements}
C. W. Tan acknowledges helpful discussions on the subject with O. T. Ng and K. Marimuthu of the Singapore National Centre for Infectious Diseases.

\appendix
\section{Appendix}

\subsection{Proof of Theorem 1}
Let $\mathbb{G}_N$ be a $d$-regular tree and $\overline{v}\in \mathbb{G}_N$ satisfies (\ref{eq:condt}), we assume that the maximum distance from $\overline{v}$ to all leaves nodes is $k+1$. Then by (\ref{eq:condt}), there are only two possible distances from $\overline{v}$ to all leaves, which are $k$ and $k+1$. For simplicity, let $T_{d,k}$ denote the perfect $d$-regular tree with $k$ levels, and we assume that the root level is $0$. We denote $t^u_v$ as the size of the subtree rooted at $v$ after removing the edge $\{(u,v)\}$  from $\mathbb{G}_N$. We have the following properties of $\mathbb{G}_N$:
\begin{enumerate}
    \item The topology of $\mathbb{G}_N$ is composed of $T_{d,k}$ plus some node $v$ that satisfies $\textup{dist}(\overline{v},v)=k+1$ and  $v$ is connected to a leaf node in $T_{d,k}$, moreover, we have $\overline{v}=v^*_N$.
    \item If $\overline{v}$ is the root of $\mathbb{G}_N$, then all subtree of $\mathbb{G}_N$ satisfy (\ref{eq:condt}).
\end{enumerate}
The proof of the above two properties can be found in the following subsections. Since $\mathbb{G}$ is a $d$-regular tree, finding the superspreader by \eqref{eq:est1} is equivalent to finding its centroids \cite{JSTSP2018}, i.e., $v^*_n$ is the centroid of $\mathbb{G}_N$. Let $\overline{v}$ be the root of $\mathbb{G}_N$, then we can prove Theorem \ref{thm:bfs} by showing that if $v^*_i\neq v^*_{i+1}$, then $v^*_{i+1}$ is the parent of $v^*_i$. To contrary,  assume $\exists i< N$ such that $v^*_i\neq v^*_{i+1}$ and $v^*_{i+1}=v_c$, where $v_c$ is a child of $v^*_i$. Without loss of generality, we can assume that $v^*_i$ is at the $l$th level of $\mathbb{G}_N$. Then, from the above property 1, we can deduce that $t_{v_c}^{v^*_{i}}>t_{v^*_{i}}^{v_c}$ since $v_c$ is the centroid of $\mathbb{G}_N$. Note that the maximum size of $t_{v_c}^{v^*_{i}}$ is $\frac{(d-1)^{k-l+1}-1}{d-2}$ since the subtree rooted at $v_c$ also satisfies (\ref{eq:condt}) by the above property 2. Similarly, since the BFS algorithm discovers all vertices at a distance $k-l$ from $v^*_i$, we have the minimum size of $t_{v^*_{i}}^{v_c}=(d-2)\frac{(d-1)^{k-l}-1}{d-2}+\frac{(d-1)^{k-l}-1}{d-2}+1$, where the first term of this equation is the sum of the size of all other subtrees rooted at $v^*_i$ except the one rooted at $v_c$. The second term is the size of $t^{v^*_{i}}_{v_p}$ where $v_p$ is the parent of $v^*_i$, and the third term is $v^*_i$ itself. Hence, we have $t_{v_c}^{v^*_{i}}=t_{v^*_{i}}^{v_c}$ which contradicts the assumption that the $v_c$ is the centroid of $\mathbb{G}_N$. We can conclude that $v^*_{i+1}$ must be the parent of $v^*_{i}$ if $v^*_i\neq v^*_{i+1}$ which implies that the trajectory of $v_{n}^*$ is the shortest path to the root $v^*_N=\overline{v}$.

\subsection{Proof of Property 1 of $\mathbb{G}_N$ used in Theorem 1}
Let $\mathbb{G}_N$ be a $d$-regular tree and $\overline{v}\in \mathbb{G}_N$ satisfies (5), we assume that the maximum distance from $\overline{v}$ to all leaves nodes is $k+1$. Then by (5), there are only two possible distances from $\overline{v}$ to all leaves, which are $k$ and $k+1$. By the definition of $d$-regular tree and (5), we can deduce that for all $v\in \mathbb{G}_N$ $\textup{dist}(v,\overline{v})<k$, $v$ is not a leaf of $\mathbb{G}_N$ which implies the degree of $v$ must be $d$. Hence, $\mathbb{G}_N$ contains a $d$-regular tree with $k$ levels denoted as $T_{d,k}$. The rest of the nodes in $\mathbb{G}_N\backslash T_{d,k}$ are leaves in $\mathbb{G}_N$ with distance $k$ or $k+1$ from the node $\overline{v}$. Since the topology of $\mathbb{G}_N$ is composed of $T_{d,k}$ and an additional leaf-level, we can conclude that $\overline{v}=v^*_N$ \cite{JSTSP2018,ShahTransIT2011}.

\subsection{Proof of Property 2 of $\mathbb{G}_N$ used in Theorem 1}
Let $\mathbb{G}_N$ be a $d$-regular tree and $\overline{v}\in \mathbb{G}_N$ satisfies (5), we assume that the maximum distance from $\overline{v}$ to all leaves nodes is $k+1$. Let $\overline{v}$ be the root of $\mathbb{G}_N$, and $v\in \mathbb{G}_N$ be an arbitrary node. If $v$ is a leaf or the root of $\mathbb{G}_N$, then the subtree rooted at $v$ satisfies (5). If $v$ is not a leaf or the root of $\mathbb{G}_N$, then we prove this property by contradiction. Assume that the subtree rooted at $v$, denoted as $T_{v}^{\overline{v}}$ does not satisfy (5), then there are two leaves $u,w \in T_{v}^{\overline{v}}$, satisfy $|\textup{dist}(u,v)-\textup{dist}(w,v)|> 1$. since $\mathbb{G}_N$ is a tree, we have $\textup{dist}(\overline{v},u)=\textup{dist}(\overline{v},v)+\textup{dist}(v,u)$ and $\textup{dist}(\overline{v},w)=\textup{dist}(\overline{v},v)+\textup{dist}(v,w)$
which implies $|\textup{dist}(\overline{v},u)-\textup{dist}(\overline{v},w)|> 1$ and contradicts to our assumption.

\subsection{Proof of Theorem 2}
We can prove this theorem by showing that if $(v^*_{j_1}=v,v^*_{j_1+1}=u)\in S_2$ and $(v^*_{j_2}=v,v^*_{j_2+1}=w)\in S_2$ where $j_2 > j_1$, then we have $j_2>2{j_1}$.
Assume there are two pairs $(u,v), (w,v)\in S_3$ where $u,v,w\in \mathbb{G}_N$. Since each pair in $S_3$ must have a corresponding pair in $S_2$, there are $(v^*_{j_1}=v,v^*_{j_1+1}=u)\in S_2$ and $(v^*_{j_2}=v,v^*_{j_2+1}=w)\in S_2$. If we treat the index case $v^1$ as the root of $\mathbb{G}_N$, then we have $v^*_{j_1}$ is the parent of $v^*_{j_1+1}$ and $v^*_{j_2+1}$. 
Let $u$ be the $d_1$th node and $w$ be the $d_2$th node visited by the DFS strategy respectively, i.e., $v^*_{j_1+1}=v^{d_1}$ and $v^*_{j_2+1}=v^{d_2}$ where $d_2>d_1$.

Note that if a node in $\mathbb{G}_N$ with weight strictly less than $n/2$, then the node is the unique ML estimator on contact tracing network $\mathbb{G}_N$ \cite{zelinka1968medians,JSTSP2018, shah2012rumor}. Hence we have $j_1+1 = 2d_1-1.$. Moreover, since  $(v=v^*_{j_2},v^*_{j_2+1})\in S_2$,  we have
\begin{align*}
 j_2+1 = 2d_2-1> 2(j_1+1)-1 = 2j_1+1.
\end{align*}
The second inequality follows by $d_2>j_1+1$ since $w$ will not be visited until the subtree $T^v_u \subseteq \mathbb{G}_N$ is fully visited by the DFS algorithm. 
We can conclude that if $v$ is involved in more that one pairs say $(u,v)$ and $(w,v)\in S_3$ then there are corresponding pairs $(v=v^*_{j_1},u=v^*_{j_1+1})$, and $(v=v^*_{j_2},w=v^*_{j_2+1})\in S_2$ in $S_2$ with index $j_2>2 j_1$. From Lemma \ref{lem:comp_S}, we have $|S_2|\leq (N-1)/2$, which implies $v$ cannot be involved in more than $\log_2 \frac{N-1}{2}$ pairs in $S_3$, since each pair $(u,v)\in S_3$ must have a corresponding $(v,u)\in S_2$.

\subsection{Proof of Lemma 1}
Since the DFS contact tracing starts from the index case $v^1$, we can assume that the index case $v^1$ is the root of the DFS tree. Now, we can distinguish parent-node and child-node in each pair $(v^*_j,v^*_{j+1})\in S$. Due to the property of the DFS tracing, a node is fully visited when all its descendants are fully visited; we have the following observation: If $(v^*_j=u,v^*_{j+1}=v)\in S_2 \cup S_3$ and $v$ is the parent of $u$, then there is no other $i>j$ such that $(v^*_i=v,v^*_{i+1}=u)\in S_2 \cup S_3$. Moreover, we can observe that: 
\begin{enumerate}
    \item If $(u,v)\in S_2$, then $u$ is the parent node of $v$.
    \item If $(u,v)\in S_3$, then $u$ is the child of $v$.
\end{enumerate}

Since the estimated superspreader changes from $u$ to $v$ only when $P(\mathbb{G}_N \mid v)>P(\mathbb{G}_N \mid u)$. We can conclude that both type two and type three transitions must be followed by a type one transition, which means the size of the type one transition is more than type two and type three. Hence, we have $|S_1|\geq |S_2|+|S_3|$. Moreover, from the above two properties, we have the fact that for each $(u,v)\in S_3$, there is a corresponding $(v,u)\in S_2$, i.e., $|S_2|\geq |S_3|$. Combining the results above, we have  $|S_1|\geq |S_2|\geq |S_3|$. Since $|S_1|+|S_2|+|S_3|=N-1$ and $|S_2|\geq |S_3|$, we have $|S_1|\geq \frac{N-1}{2}$ and $|S_3| \leq \frac{N-1}{4}$.

\subsection{Proof of Theorem 3}
As shown in \eqref{eq:est1}, solving the MLE problem needs the information of all the permitted permutations for all the nodes in $G_{n}$. Since a newly infected node can only be infected by one of its infected neighbors, $v_{i+1}$ must be the neighbor of $v_{i}$ in any permitted permutation $\sigma$, and thus calculating $p(\sigma \mid v)$ with \eqref{eq:prob2} is a recursive process of aggregating the information of degree $d(v_l)$ and $\Phi_l$ from the $l$th layer's neighbors of node $v$ in $\sigma$, which can be expressed as
 \begin{equation}
\mathbb{P}^{(l)}(\sigma \mid v) = \sum_{u\in N_G(v)}f\big(\mathbb{P}^{(l-1)}(\sigma \mid v),\mathbb{P}^{(l-1)}(\sigma \mid u)\big),\label{eq:agg1}
\end{equation}
with $\mathbb{P}^{(l)}(\sigma \mid v)|_{l=1} = 1/d(v)$. Therefore, it is a recursive iteration to calculate $\sum_{\sigma \in \Omega(G_{n} \mid v)} \mathbb{P}(\sigma | v)$ to aggregate the information of degree $d(v_l)$ and $\Phi_l$ from its $l$th layer's neighbors in all $\sigma\in \Omega(G_{n} \mid v)$, which can be shown as
 \begin{align}\label{eq:agg2}
&\;\;\;\;\;\mathbb{P}^{(l)}(G_{n} \mid v) \\\nonumber
&=\sum_{\sigma \in \Omega(G_{n} \mid v)} \sum_{u\in N_G(v)}f\big(\mathbb{P}^{(l-1)}(\sigma | v),\mathbb{P}^{(l-1)}(\sigma | u)\big)\\\nonumber
&=\sum_{u\in N_G(v)}g\big(\mathbb{P}^{(l-1)}(G_{n} | v),\mathbb{P}^{(l-1)}(G_{n} | u)\big).
\end{align}
Note that \eqref{eq:agg2} demonstrates the process of iterative aggregation to calculate $\mathbb{P}(G_{n} \mid v)$ for node $v$. Since $\mathbb{G}$ is connected, the shortest distance between any two nodes satisfies $\textup{dist}(u,v)\leq \textup{diam}(G)$. Thus, after some $\textup{diam}(G)$ steps of aggregation and combination, each node can obtain the feature information of all nodes in the graph. Furthermore, it is evident that both $f(\cdot)$ and $g(\cdot)$ are continuous functions. Consequently, based on the universal approximation theorem, these functions can be approximated to any desired degree of precision using standard multilayer perceptrons with hidden layers and non-constant, monotonically increasing activation functions \cite{hornik1991approximation}.

\bibliographystyle{IEEEtran}
\bibliography{main}

\vspace{-4em}
\begin{IEEEbiography}[{\includegraphics[width=1in,height=1.25in,clip,keepaspectratio]{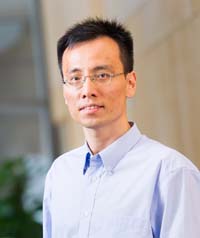}}]
{Chee Wei Tan}(M '08-SM '12)  received the M.A. and Ph.D. degrees in electrical engineering from Princeton University, Princeton, NJ, USA, in
2006 and 2008, respectively. He is an Associate Professor of Computer Science with Nanyang Technological University, Singapore. Dr. Tan has served as the Editor for IEEE Transactions on Cognitive Communications and Networking, IEEE/ACM Transactions on Networking, IEEE Transactions on Communications and as IEEE ComSoc Distinguished Lecturer. He received the Princeton University Wu Prize for Excellence, Google Faculty Award, teaching excellence awards, and was selected twice for the U.S. National Academy of Engineering China-America Frontiers of Engineering Symposium. His research interests are in networks, distributed optimization and Generative AI. 
\end{IEEEbiography}

\vspace{-4em}
\begin{IEEEbiography}[{\includegraphics[width=1in,height=1.25in,clip,keepaspectratio]{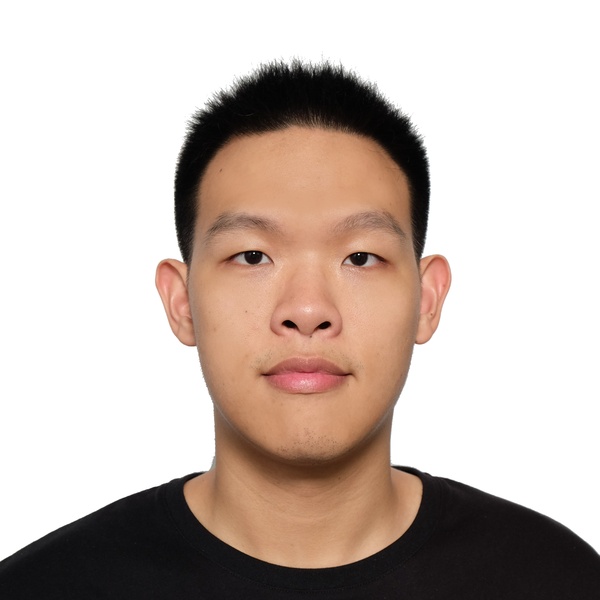}}]
{Pei-Duo Yu}  received the B.Sc. and M.Sc. degree in Applied Mathematics in 2011 and 2014, respectively, from the National Chiao Tung University, Taiwan. He received his  Ph.D. degree at the Department of Computer Science, City University of Hong Kong, Hong Kong. Currently, he is an Assistant Professor at the Department of Applied Mathematics in Chung Yuan Christian University, Taiwan. His research interests include combinatorics counting, graph algorithms, optimization theory and its applications.  
\end{IEEEbiography}

\vspace{-4em}
\begin{IEEEbiography}[{\includegraphics[width=1in,height=1.25in,clip,keepaspectratio]{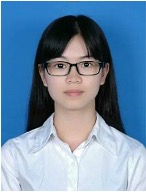}}]
{Siya Chen} received her B.Sc. and M.Sc. degrees in Applied Mathematics from Shenzhen University. She is currently working toward her Ph.D. degree in the Department of Computer Science at City University of Hong Kong. Her research interests include machine learning and optimization theory in wireless network optimization.
\end{IEEEbiography}

\vspace{-4em}
\begin{IEEEbiography}[{\includegraphics[width=1in,height=1.25in,clip,keepaspectratio]{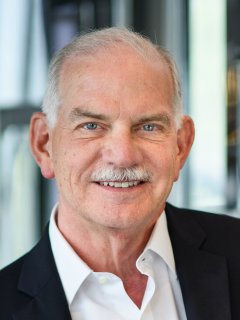}}]
{H. Vincent Poor} (S'72, M'77, SM'82, F'87) received the Ph.D. degree in EECS from Princeton University in 1977.  From 1977 until 1990, he was on the faculty of the University of Illinois at Urbana-Champaign. Since 1990 he has been on the faculty at Princeton, where he is currently the Michael Henry Strater University Professor. During 2006 to 2016, he served as the dean of Princeton's School of Engineering and Applied Science, and he has also held visiting appointments at several other universities, including most recently at Berkeley and Caltech. His research interests are in the areas of information theory, machine learning and network science, and their applications in wireless networks, energy systems and related fields. Among his publications in these areas is the book Machine Learning and Wireless Communications.  (Cambridge University Press, 2022). Dr. Poor is a member of the National Academy of Engineering and the National Academy of Sciences and is a foreign member of the Royal Society and other national and international academies. He received the IEEE Alexander Graham Bell Medal in 2017.
\end{IEEEbiography}

\end{document}